\documentclass[11pt, a4paper, copyright, gdm]{google}

\usepackage[authoryear, sort&compress, round]{natbib}

\usepackage{autobreak}
\usepackage{nccmath}

\usepackage{multirow}
\usepackage{float}
\usepackage{subcaption}   
\usepackage{graphbox}
\usepackage{wrapfig}
\usepackage{rotating}
\usepackage{tablefootnote}
\usepackage{nicematrix}
\usepackage{tikz}
\usetikzlibrary{fit,calc,arrows.meta,positioning,backgrounds}

\usepackage{setspace}
\usepackage{xspace}
\usepackage{csquotes}
\usepackage{xurl}
\usepackage[ruled,linesnumbered]{algorithm2e}
\SetKwInput{Input}{Input}
\SetKwInput{Output}{Output}
\SetKwBlock{Begin}{begin}{end}

\usepackage{gdefs}   
\usepackage{refs}    
\newtheorem*{theorem*}{Theorem}

\newcommand{\Omit}[1]{}

\newcommand{\mas}[1]{{#1}}
\newcommand{\asb}[1]{{#1}}

\newif\ifarxiv \arxivtrue
\newcommand{\sys}{\texorpdfstring{\texttt{Tether}}{Tether}\xspace}
\newcommand{\isolate}{\texorpdfstring{\texttt{Isolate}}{Isolate}\xspace}


\uselogo{} 

\title{\sys: Verified Stateful Policy Engine for System-Level Monitoring of Agents}

\newcommand{\aditya}[1]{}
\newcommand{\revcolor}{}

\author[1]{Meghana Aparna Sistla}
\author[1,3]{Alaia Solko-Breslin}
\author[2,4]{Somesh Jha}
\author[2]{Aditya Kanade}
\author[1]{Krishnamurthy Dj Dvijotham}

\affil[1]{\thepa{}{}}
\affil[2]{Google}
\affil[3]{University of Pennsylvania}
\affil[4]{University of Wisconsin-Madison}

\begin{abstract}
Autonomous AI agents are increasingly deployed for long-running tasks that execute arbitrary shell commands or programs. However, these actions can result in unauthorized network requests or persistent, malicious changes to the user's file system, processes, or privileges. Existing defense mechanisms rely on either LLM-based monitors, which lack provable safety guarantees, or stateless policy enforcement layers, which miss multi-step data flows and execution side-effects unmodeled by the policy language (e.g., network calls initiated inside a Python script). We introduce \sys, a framework for runtime enforcement of stateful security policies over the observed side-effects of an agent's actions. \sys executes each proposed action in an ephemeral, isolated sandbox, inferring abstract action semantics from observed system calls. \sys then enforces stateful policies over these inferred semantics, permitting actions to proceed only if they comply with the policy. Policies and action semantics are defined in an embedded DSL in Lean 4 capable of expressing diverse constraints, including information flow control (IFC), access control, and network security policies. Because actions can propagate labels from sources to destinations, a single rule can enforce fine-grained IFC policies across multi-step execution flows of arbitrary length. Expressing policies in Lean enables formal verification that they satisfy global security invariants (e.g., preventing network access after reading a sensitive file). We evaluate \sys on realistic coding-agent benchmarks across both benign and adversarial trajectories. With a single policy shared across all tasks, \sys blocks 100\% of in-scope attacks while maintaining a benign false-positive rate under 5\%.
\end{abstract}

\begin{document}

\maketitle

\section{Introduction}
\label{Se:intro}

Large Language Model (LLM) powered agents are increasingly deployed to automate complex, multi-step workflows~\citep{yao2023react, swe-agent, llm-agent-survey}.
While these agents are clearly useful, they also introduce critical security risks.
Agents can, due to misalignment or due to a prompt injection,  exfiltrate sensitive data or download and execute malicious binaries \citep{greshake2023youvesignedforcompromising, zhan-etal-2024-injecagent, debenedetti2024agentdojo, ruan2024identifying}.

To mitigate these risks, many production agent harnesses deploy a secondary LLM-as-a-judge monitor to review proposed commands before execution~\citep{anthropic2026automode, openai2026autoreview}.
While auxiliary LLMs can assess high-level intent, they remain probabilistic and susceptible to prompt injections~\citep{remedios2026redteamingautomode}, rendering them incapable of providing formal security guarantees.
An alternative is to use formal reference monitors~\citep{Anderson:1972} integrated into sandboxes that enforce explicit security policies over agent actions~\citep{nvidia2026openshell, aws2026dogwood, aws2026strandsbox, wu-isolategpt}.
Enforcing security policies on autonomous terminal agents presents two fundamental challenges unresolved by prior work:

\textbf{Challenge 1: Ambiguous tool semantics.}
Existing reference monitors for agents assume that actions taken by agents are structured API calls whose semantics are known and targets are explicit in their arguments. They may also attempt to filter shell commands statically before execution~\citep{cutler2024cedar, shi2026progent}.
In a terminal environment, however, a single command (e.g., running a Python script or a shell pipeline with variable expansion) can  perform file and network operations that are difficult to predict statically~\citep{greenberg2020smoosh}.
Conversely, simply executing the command to observe its side effects may mutate the host system or leak sensitive data over the network before the monitor has the opportunity to block it.

\textbf{Challenge 2: Formally verified, stateful IFC.}
To prevent multistage attacks, such as staging a secret across temporary files before uploading it, a runtime monitor must enforce stateful Information Flow Control (IFC)~\citep{denning1976latticeifc, certification-programs-secure-ifc, decentralized-model-ifc}.
However, existing frameworks like Cedar~\citep{cutler2024cedar} and OpenShell~\citep{nvidia2026openshell} rely on stateless allowlists, while event-history monitors~\citep{aws2026dogwood, aws2026strandsbox} maintain no persistent security labels on resources, forcing policies to reconstruct flows over a fixed number of hops (see~\sectref{related_work}).
Furthermore, as policies grow to combine access control, multi-dimensional taint tracking, and policy composition, policy authors need formal guarantees that their rules are free of subtle bypasses.

We introduce \sys, a stateful runtime reference monitor and verified policy engine for terminal agents.
Rather than assuming what a shell command might do from its syntax, \sys executes each command inside an ephemeral copy-on-write (CoW) filesystem layer~\citep{lamprou2026try} under system-call tracing, infers the command's semantics, and enforces formal security policies.
To infer a command's semantics safely, \sys traces system calls in the CoW layer and intercepts outbound network requests via a local proxy. The filesystem changes and network requests are committed (or forwarded) if and only if the policy permits every action seen so far; otherwise, the ephemeral layer and intercepted requests are cleanly discarded.

From the observed system calls, \sys infers a sequence of canonical abstract actions, e.g., \act{read}, \act{copy}, and \act{uploadNet}.
Because reconstructing the exact data flow of an arbitrary process from system calls alone is impossible in general, \sys does not assume that the exact semantics of any terminal action can be recovered; instead, it over-approximates the command's information-flow and privilege side effects.
For example, if a script reads from file \texttt{a.txt} and subsequently writes to file \texttt{b.txt}, \sys conservatively infers a data-flow transfer from \texttt{a.txt} to \texttt{b.txt} even if the written bytes were independent of \texttt{a.txt}.
This over-approximation may introduce occasional false positives, but avoids missing tracked data flows (\sectref{limitations}).

We implement our abstract action calculus and policy language as an embedded domain-specific language (EDSL) in Lean~4~\citep{lean4}, a dependently typed programming language and interactive theorem prover with a small trusted kernel\footnote{An EDSL uses the host language's parser and compiler. In Lean, this ensures that the verified policy compiles directly to the runtime monitor.}.
Each abstract action deterministically updates a security state parametric over an arbitrary product lattice of security labels (e.g., confidentiality and integrity), propagating resource labels across direct transfers (e.g., \act{copy} or \act{concatenate}) without over-tainting the agent's context.
To aid policy designers, we mechanically prove key metatheoretic guarantees in Lean~4 (\sectref{dsl-metatheory}), including \emph{deny-overrides monotonicity}, \emph{non-bypassability of required policies}, \emph{permutation invariance} (order independence), and \emph{state preservation under denial}.
Because \sys decouples semantic inference (our \isolate engine, \sectref{isolate}) from policy specification and enforcement, the inference engine can be improved independently (\sectref{related_work}).

We evaluate \sys on two realistic coding-agent benchmarks that pair benign software-engineering tasks with adversarial side tasks: MonitoringBenchHonest~\citep{kucukkurt2026monitoringbenchhonest} and LinuxArena~\citep{linuxarena}.
Using a \emph{single} policy parameterized only by task-specific initial sensitivity labels, we evaluate \sys across $199$ adversarial and $218$ benign trajectories on MonitoringBenchHonest and $145$ adversarial and $1{,}372$ benign trajectories on LinuxArena.
On all in-scope trajectories (where all actions are traced by our system), \sys achieves $100\%$ adversarial recall with a low false-positive rate of $2.3\%$ on MonitoringBenchHonest and $5.0\%$ on LinuxArena.
The out-of-scope trajectories (\sectref{limitations}) comprise either (i)~interactions with untraced external state (such as background daemons, database queries, etc.), which cannot be tracked at a fine grain but which \sys can deterministically detect and block at the boundary, or (ii)~purely semantic attacks that violate no system-level flow invariants and thus cannot be prevented by any formal guardrail without sacrificing utility.

In summary, this paper's contributions are as follows:
\begin{enumerate}
  \item We develop a compositional, lattice-parametric policy EDSL in Lean~4 defined over an abstract action calculus that unifies access control and IFC (\sectref{lean_dsl}).
  \item We show that inferring and mapping the semantics of concrete actions to abstract actions helps enforce fine-grained policies. For coding agents, we show that system-call tracing (paired with copy-on-write sandboxing) is a useful approach to inferring the semantics of actions, and use fuzzing to validate this semantic inference engine (\sectref{fuzzer}).
  \item We provide a mechanism for specifying user-intended invariants and proving that the policy satisfies them (\sectref{verify-policy}).
  \item We present, to the best of our knowledge, the first evaluation of a formal runtime monitor on realistic coding-agent benchmarks: MonitoringBenchHonest and LinuxArena (\sectref{experiments}), blocking $100\%$ of adversarial trajectories that can be modeled by \sys, with low false-positive rates ($2.3\%$--$5.0\%$).
\end{enumerate}

\section{Overview}
\label{Se:overview}

\subsection{Motivating Example}
\label{Se:running-example}

Consider an agent that has access to a sensitive file, \file{data/servers.csv}, containing a company's inventory.
The user wants to prevent this file from being sent to an external site (\texttt{api.github.com}), but also allow the agent to read it for local checks and upload benign status reports (\file{scratch/report.txt}) to the same site.
The agent proposes the following shell commands (\lstref{running}):

\smallskip\noindent\begin{minipage}{\columnwidth}
\ifarxiv\centering\captionsetup{justification=centering}\begin{minipage}{20.5em}\fi
{\footnotesize
\begin{verbatim}
$ curl --json @scratch/report.txt \
    https://api.github.com         # (1)
$ cp data/servers.csv \
    scratch/report.txt             # (2)
$ curl --json @scratch/report.txt \
    https://api.github.com         # (3)
\end{verbatim}
}
\ifarxiv\end{minipage}\fi
\captionof{listing}{The running example.}
\label{Lst:running}
\end{minipage}\par\smallskip

\noindent Step~(1) uploads the benign report,
step~(2) copies the sensitive inventory to \file{scratch/report.txt}, and
step~(3) uploads the report again.
Although Steps~(1) and~(3) run the same command, a runtime monitor must allow Step~(1) and deny Step~(3) because the report is now sensitive.

Unlike existing frameworks that rely on stateless allowlists, lack persistent resource state, or coarsely approximate command effects~\citep{nvidia2026openshell, aws2026dogwood, aws2026strandsbox, cedar-verification-guided-approach}, \sys dynamically traces and evaluates arbitrary command execution at runtime.
By tracing low-level system calls, \sys infers what a command actually does, maps it to a calculus of abstract actions (e.g., \act{copy} or \act{uploadNet}), and checks it against a stateful policy.
To ensure security, commands run speculatively; \sys commits their environmental side effects if and only if the policy approves the inferred actions.



To demonstrate how \sys distinguishes between these steps, we define a policy over a security lattice where $\sigma(r, \mathit{key})$ denotes the security label of resource $r$'s $\mathit{key}$ attribute. 
We initialize the labels such that $\sigma($\file{data/servers.csv}$,\allowbreak\ \pol{content}) = \lbl{Sensitive}$ and all other files to $\lbl{Public}$ (e.g., $\sigma($\file{scratch/report.txt}$,\allowbreak\ \pol{content}) = \lbl{Public}$), with $\lbl{Public} \sqsubseteq \lbl{Sensitive}$. 
The following is a simplified version of our policy that permits uploads only for public files\footnote{The complete policy is augmented with rules for agent's content taint, direct network sends, etc. to cover all possible cases.}.

\begin{center}
\small\polsf
\begin{tabular}{@{}l@{}}
\kw{rule} restrict\_upload\_exfil :\\
\hspace*{1em}\kw{on} uploadNet($\mathit{src}$, $\mathit{dest}$)\\
\hspace*{1em}\kw{if} $\sigma$[$\mathit{src}$, content] $\sqsubseteq$ \lbl{Public}\\
\hspace*{1em}\kw{then} \kw{allow} \kw{else} \kw{deny}
\end{tabular}
\end{center}

When an action is approved, \sys dynamically updates the affected resource labels according to the action's formal semantics (see~\sectref{lean_dsl}).
To infer abstract actions from a shell command, \sys traces its system calls and maps them to high-level operations.

\par\smallskip
\noindent\textbf{Step~(1):} \texttt{curl} on \file{scratch/report.txt} invokes \texttt{openat}, \texttt{read}, \texttt{connect}, and \texttt{sendto}, which \sys maps to \act{uploadNet}(\file{scratch/report.txt},\allowbreak\ \file{api.github.com}). Since $\sigma($\file{scratch/report.txt}$,\allowbreak\ \pol{content}) = \lbl{Public} \sqsubseteq \lbl{Public}$, \sys allows the action.\par\smallskip

\noindent\textbf{Step~(2):} \texttt{cp} from \file{data/servers.csv} to \file{scratch/report.txt} invokes \texttt{openat}, \texttt{read}, and \texttt{write}, which \sys maps to \act{copy}(\file{data/servers.csv},\allowbreak\ \file{scratch/report.txt}). With no matching rule, the action is allowed by default, updating $\sigma($\file{scratch/report.txt}$,\allowbreak\ \pol{content})$ to \lbl{Sensitive}.\par\smallskip

\noindent\textbf{Step~(3):} Running the same \texttt{curl} command from Step~(1) produces an identical \act{uploadNet} action, which \sys denies because $\sigma($\file{scratch/report.txt}$,\allowbreak\ \pol{content}) = \lbl{Sensitive} \not\sqsubseteq \lbl{Public}$.\par\smallskip

Because \sys tracks information flow dynamically, the same policy generalizes to multi-hop flows. 
For instance, if the agent stages the sensitive inventory through intermediate temporary files (Steps~(2a)--(2b) in the following listing), \sys propagates the \lbl{Sensitive} label across each copy, blocking the subsequent upload in Step~(3). 
Conversely, copies and uploads of independent public files (Steps~(4)--(5)) are allowed, avoiding false positives.

\smallskip\noindent\begin{minipage}{\columnwidth}
\ifarxiv\centering\begin{minipage}{25em}\fi
{\footnotesize
\begin{verbatim}
$ cp data/servers.csv scratch/tmp.csv       # (2a)
$ cp scratch/tmp.csv scratch/report.txt     # (2b)
$ curl --json @scratch/report.txt \
    https://api.github.com                   # (3)
$ cp templates/status.txt scratch/status.txt # (4)
$ curl --json @scratch/status.txt \
    https://api.github.com                   # (5)
\end{verbatim}
}
\ifarxiv\end{minipage}\fi
\end{minipage}\ifarxiv\par\smallskip\fi

\subsection{Architecture of \sys}
\label{Se:architecture-overview}
\sys consists of two main components: \isolate (\sectref{isolate}), a semantic inference engine that sandboxes agent execution to infer abstract actions, and a stateful policy monitor implemented in Lean~4 (\sectref{lean_dsl}).
The mapping from low-level observed effects to canonical abstract actions serves as our \emph{semantic contract}, modulo which \sys's security guarantees hold.

Building on the filesystem-isolation model in \texttt{try}~\citep{lamprou2026try}, \isolate runs each proposed command inside a disposable OverlayFS layer under system-call tracing. It intercepts system-level events, gates outbound network requests through a local proxy, and tracks data provenance to emit a sequence of abstract actions (e.g., \act{copy} or \act{uploadNet}).
These sequences of abstract actions are then evaluated by the Lean~4 reference monitor. The monitor tracks the system state such as every resource's security labels ($\sigma$) and the agent's context taint, thereby allowing or denying actions based on the specified policy, while updating the system state accordingly.
The staged changes are committed or discarded based on the monitor's verdict.

\subsection{Threat Model}

We assume that \sys operates within the reference monitor framework \citep{Anderson:1972}, enforcing deterministic security policies over an untrusted terminal agent.
We assume that an adversary can control the agent's actions either directly (as a malicious user) or indirectly via prompt-injection attacks.
Because we treat the agent's underlying LLM as untrusted, we assume that the adversary can generate arbitrary terminal commands.
The adversary has full knowledge of the policy but cannot bypass or alter the monitor or its policy rules.

As is standard in the sandboxing literature, our Trusted Computing Base (TCB) consists of the host OS kernel, the system-call tracing layer, the ephemeral filesystem overlay driver, the network proxy, \sys's abstract action inference, and the Lean~4 proof kernel (for verification) alongside the Lean~4 compiler and runtime.
Kernel-level exploits and sandbox escapes are outside the scope of this work.

\section{Policy DSL and Operational Semantics}
\label{Se:lean_dsl}

This section presents the policy language and operational semantics of \sys. We implement the language as an embedded domain-specific language (EDSL) in Lean~4, so a single definition serves as both the runtime monitor and the formal specification for mechanized proofs\footnote{In our prototype, the Lean~4 evaluator executes policies directly. Higher-throughput deployments can pair the Lean~4 specification with a compiled runtime in a systems language such as Rust and validate equivalence via differential testing, as in Cedar~\citep{cedar-verification-guided-approach}.}.

\subsection{Security Lattice}
\label{Se:dsl-lattice}

\sys is parametric over the space of security labels. 
All abstract actions, policies, and proofs are defined on a bounded lattice $(L, \sqsubseteq, \sqcup, \sqcap, \bot, \top)$, where $\sqsubseteq$ is a partial order over labels, $\sqcup$ is the least upper bound (join), $\sqcap$ is the greatest lower bound (meet), $\bot$ is the least element (least restrictive label), and $\top$ is the greatest element (most restrictive label). Every label $\ell \in L$ satisfies $\bot \sqsubseteq \ell \sqsubseteq \top$, and the join $\ell_1 \sqcup \ell_2$ is the least upper bound of $\ell_1$ and $\ell_2$:
\begin{gather*}
  \ell_1 \sqsubseteq \ell_1 \sqcup \ell_2, \qquad
  \ell_2 \sqsubseteq \ell_1 \sqcup \ell_2, \\
  \ell_1 \sqsubseteq \ell \wedge \ell_2 \sqsubseteq \ell \implies \ell_1 \sqcup \ell_2 \sqsubseteq \ell.
\end{gather*}

We can represent multiple security dimensions (e.g., confidentiality and integrity) corresponding to component lattices $L_1, \dots, L_n$ with a product lattice $L = L_1 \times \dots \times L_n$ with component-wise ordering, join, and bounds:
{\medmuskip=2mu
\begin{gather*}
  (a_1, \dots, a_n) \sqsubseteq (b_1, \dots, b_n) \iff \forall i,\; a_i \sqsubseteq b_i, \\
  (a_1, \dots, a_n) \sqcup (b_1, \dots, b_n) = (a_1 \sqcup b_1, \dots, a_n \sqcup b_n),
\end{gather*}}
Because \sys is lattice-generic, adding new dimensions does not affect the transition rules or proofs.

\subsection{Security State and Action Transitions}
\label{Se:dsl-transitions}

We model \sys's security semantics by defining an abstract execution state over a set of core domains, and formalizing how individual agent actions transition this state.
We first define the three core system domains.

\begin{enumerate}
  \item \textbf{Resources} ($\mathit{Resource}$): A resource identifier is a canonical representation of a filesystem path (e.g., \texttt{/etc/shadow}, \texttt{data/servers.csv}), network endpoint (e.g., \texttt{api.github.com}, \texttt{postgres}), or synthetic system domain (e.g., \texttt{<system:binaries>}).
  \item \textbf{Attribute keys} ($\mathit{Key}$): Every resource has attributes such as its content, location, and size, i.e., \pol{content}, \pol{location}, \pol{size\_bytes}.
  \item \textbf{Privileges} ($\mathit{Privilege}$): These represent agent privilege levels, i.e., $\mathit{Privilege} = \{\lbl{User}, \lbl{Admin}\}$.
\end{enumerate}

\begin{figure*}[tb!]
\centering
\small
\renewcommand{\arraystretch}{1.3}
\begin{tabular}{@{}lll@{}}
\toprule
\textbf{Action} $a$ & \textbf{Updated State} $S' = \delta(\langle \mathcal{T}, \mathcal{P}, \sigma \rangle, a)$ & \textbf{Rule Name} \\
\midrule
$\act{read}(r, k)$ & $\langle \mathcal{T} \sqcup \sigma(r, k),\; \mathcal{P},\; \sigma \rangle$ & \textsc{E-Read} \\
$\act{recvNet}(\mathit{src})$ & $\langle \mathcal{T} \sqcup \sigma(\mathit{src}, \pol{content}),\; \mathcal{P},\; \sigma \rangle$ & \textsc{E-Recv} \\
$\act{write}(r, k, \cdot, \cdot)$ & $\langle \mathcal{T},\; \mathcal{P},\; \sigma[(r, k) \mapsto \sigma(r, k) \sqcup \mathcal{T}] \rangle$ & \textsc{E-Write} \\
$\act{create}(\mathit{srcs}, r, \cdot, \cdot)$ & \begin{tabular}[t]{@{}l@{}}$\langle \mathcal{T},\; \mathcal{P},\; \sigma[(r, \pol{content}) \mapsto \mathcal{T},$\\ $\quad (r, k) \mapsto \mathcal{T} \sqcup \operatorname{loc}(r) \sqcup \bigsqcup_{s \in \mathit{srcs}} \sigma(s, \pol{content}) \mid k \neq \pol{content}] \rangle$\end{tabular} & \textsc{E-Create} \\
$\act{delete}(r, \cdot, \cdot)$ & $\langle \mathcal{T},\; \mathcal{P},\; \sigma[(r, k) \mapsto \bot \mid k \in \mathit{Key}] \rangle$ & \textsc{E-Delete} \\
$\act{copy}(\mathit{src}, d, \cdot, \cdot)$ & $\langle \mathcal{T},\; \mathcal{P},\; \sigma[(d, \pol{content}) \mapsto \sigma(\mathit{src}, \pol{content}) \sqcup \operatorname{loc}(d)] \rangle$ & \textsc{E-Copy} \\
$\act{concatenate}(\mathit{srcs}, d, \cdot, \cdot)$ & $\langle \mathcal{T},\; \mathcal{P},\; \sigma[(d, \pol{content}) \mapsto (\bigsqcup_{s \in \mathit{srcs}} \sigma(s, \pol{content})) \sqcup \operatorname{loc}(d)] \rangle$ & \textsc{E-Concat} \\
$\act{downloadNet}(\mathit{src}, d)$ & $\langle \mathcal{T},\; \mathcal{P},\; \sigma[(d, \pol{content}) \mapsto \sigma(\mathit{src}, \pol{content}) \sqcup \operatorname{loc}(d)] \rangle$ & \textsc{E-Download} \\
$\act{sendNet}(\mathit{dest}) \mid \act{uploadNet}(\mathit{src}, \mathit{dest})$ & $\langle \mathcal{T},\; \mathcal{P},\; \sigma \rangle$ & \textsc{E-Egress} \\
$\act{escalatePrivilege}(p, \cdot)$ & $\langle \mathcal{T},\; p,\; \sigma \rangle$ & \textsc{E-Escalate} \\
\bottomrule
\end{tabular}
\caption{Abstract action state transition semantics $\delta(S, a) = S'$ over security state $S = \langle \mathcal{T}, \mathcal{P}, \sigma \rangle$, where $\operatorname{loc}(d) = \sigma(\operatorname{parent}(d), \pol{location})$. Arguments unused in the state transition are denoted by $\cdot$.}
\label{Fi:action-transitions}
\end{figure*}

The abstract system state is represented as a 3-tuple of the context taint $\mathcal{T}$, agent privilege level $\mathcal{P}$, and sensitivity map $\sigma$, parametric over the security lattice $L$:
\[
  S = \langle \mathcal{T},\; \mathcal{P},\; \sigma \rangle.
\]
\begin{enumerate}
  \item \textbf{Context taint} ($\mathcal{T} \in L$): The current label of the agent's working memory (initially $\bot$).
  \item \textbf{Agent privilege} ($\mathcal{P} \in \mathit{Privilege}$): This is the active privilege level of the agent (initially $\lbl{User}$).
  \item \textbf{Sensitivity map} ($\sigma : \mathit{Resource} \times \mathit{Key} \to L$): This maps each resource and attribute key to its security label $\sigma(r, k)$ and is parametric over $L$. When the key is \pol{content}, we write $\sigma(r)$ for $\sigma(r, \pol{content})$.

  \smallskip\noindent
  Given a set of user-specified sensitivity label rules
  $\mathcal{C} = \{(\mathit{pat}_i, k_i, \ell_i)\}$, where $\mathit{pat}_i$ is a pattern over resources, $k_i \in \mathit{Key}$ is a target attribute key (e.g., $\pol{content}$ or $\pol{location}$), $\ell_i \in L$ is a target security label, and $\ell_{\text{default}} \in L$ is a default label,
  let $M(r, k) = \{\ell \mid (\mathit{pat}, k, \ell) \in \mathcal{C},\; r \models \mathit{pat}\}$ be the set of labels from all rules matching resource $r$ and key $k$ (where $r \models \mathit{pat}$ denotes that $r$ matches $\mathit{pat}$). The initial map $\sigma_0$ is defined as
  \[
    \sigma_0(r, k) =
    \begin{cases}
      \bigsqcup_{\ell \in M(r, k)} \ell & \text{if } M(r, k) \neq \emptyset, \\
      \ell_{\text{default}} & \text{otherwise.}
    \end{cases}
  \]
\end{enumerate}

With this, we define 11 abstract actions and their operational semantics (\figref{action-transitions}).
The transition function $\delta(S, a) = S'$ defines how each abstract action $a$ updates the abstract system state $S = \langle \mathcal{T}, \mathcal{P}, \sigma \rangle$.
Note that $\operatorname{parent}(d)$ strips the final \texttt{/}-delimited segment of path $d$, with $\operatorname{parent}(\texttt{/}) = \texttt{/}$, and $\operatorname{loc}(d) = \sigma(\operatorname{parent}(d), \pol{location})$ denotes the parent directory's location label; non-filepaths are unchanged.
We now describe the abstract actions in more detail.

\smallskip\noindent
\textbf{Context-tainting read} (\textsc{E-Read}, \textsc{E-Recv}).
  \begin{itemize}
    \item $\act{read}(r, k)$ takes a source resource $r$ and an attribute key $k \in \mathit{Key}$ (e.g., $\pol{content}$ when reading a file, or $\pol{location}$ and $\pol{size\_bytes}$ on \texttt{stat}). The observed data enters the agent's context window, so the new context taint becomes $\mathcal{T} \sqcup \sigma(r, k)$, while $\sigma$ is unchanged.
    \item $\act{recvNet}(\mathit{src})$ occurs when a network endpoint's data is read by the agent, entering its context and updating the context taint to $\mathcal{T} \sqcup \sigma(\mathit{src}, \pol{content})$.
  \end{itemize}

\smallskip\noindent
\textbf{Context-tainted mutation} (\textsc{E-Write}, \textsc{E-Create}, \textsc{E-Delete}).
  \begin{itemize}
    \item $\act{write}(r, k, \mathit{actor}, \mathit{origin})$ writes to resource $r$ with attribute key $k \in \mathit{Key}$. 
    It also carries optional parameters $\mathit{actor}$ and $\mathit{origin}$, indicating the immediate process and originating script that triggered the write (likewise carried by \act{create}, \act{delete}, \act{copy}, and \act{concatenate}). 
    Because the written data may be derived from the agent's context, the resource's security label is updated to $\sigma(r, k) \sqcup \mathcal{T}$.
    The $\mathit{actor}$ and $\mathit{origin}$ parameters are used only by policy guards and do not affect the state transition.
    \item $\act{create}(\mathit{srcs}, r, \mathit{actor}, \mathit{origin})$ creates a new resource $r$ whose metadata may come from the resources in $\mathit{srcs}$. It sets the content label $\sigma'(r, \pol{content}) = \mathcal{T}$ and the metadata labels $\sigma'(r, k) = \mathcal{T} \sqcup \operatorname{loc}(r) \sqcup \bigsqcup_{s \in \mathit{srcs}} \sigma(s, \pol{content})$ for $k \neq \pol{content}$.
    \item $\act{delete}(r, \mathit{actor}, \mathit{origin})$ deletes resource $r$ and resets $\sigma'(r, k) = \bot$ for all $k \in \mathit{Key}$.
  \end{itemize}

\smallskip\noindent
\textbf{Storage transfer} (\textsc{E-Copy}, \textsc{E-Concat}, \textsc{E-Download}).
  \begin{itemize}
    \item $\act{copy}(\mathit{src}, d, \mathit{actor}, \mathit{origin})$ copies the source resource $\mathit{src}$ to the destination location $d$.
    \item $\act{concatenate}(\mathit{srcs}, d, \mathit{actor}, \mathit{origin})$ combines a list of source files $\mathit{srcs}$ into a destination location $d$.
    \item $\act{downloadNet}(\mathit{src}, d)$ writes a response from network endpoint $\mathit{src}$ directly to resource $d$ (e.g., \texttt{curl -o}).
    Unlike \act{read} and \act{recvNet}, these actions propagate the source content label(s) (joined with $\operatorname{loc}(d)$) directly to $\sigma'(d, \pol{content})$ without tainting the agent's context.
  \end{itemize}

\smallskip\noindent
\textbf{Network egress and privilege escalation} (\textsc{E-Egress}, \textsc{E-Escalate}).
  \begin{itemize}
    \item $\act{sendNet}(\mathit{dest})$ occurs when the agent sends data obtained from its context to network endpoint $\mathit{dest}$, and $\act{uploadNet}(\mathit{src}, \mathit{dest})$ uploads resource $\mathit{src}$ to network endpoint $\mathit{dest}$. Both leave $S$ unchanged; policies guard these actions by checking $\mathcal{T}$ and $\sigma(\mathit{src}, \pol{content})$ before execution.
    \item $\act{escalatePrivilege}(p, b)$ updates $\mathcal{P}$ to privilege level $p \in \mathit{Privilege}$, optionally recording the target binary $b$, such as a command passed to \texttt{sudo}.
  \end{itemize}

\noindent Our benchmark policy (\sectref{experiments}) is defined over these abstract actions. We did not include other standard IFC actions such as \act{declassify} and \act{endorse}~\citep{decentralized-model-ifc} only because our policy did not need them; this is not a limitation of our formalism, and they can be easily added.

\subsection{Policy DSL}
\label{Se:dsl-syntax}

We now define the syntax of the policy language, implemented as an embedded domain-specific language (EDSL) in Lean~4. A policy $\Pi$ written in the grammar shown in \figref{policy-grammar} specifies which actions are permitted in state $S$, while \figref{policy-example} shows a sample information-flow policy.



\begin{figure}[tb!]
  \centering
  \footnotesize
  \renewcommand{\arraystretch}{1.2}
  \begin{tabular}{@{}r@{\;\;}r@{\;}c@{\;}l@{}}
    \textbf{Lattices} & $\Lambda$ & $::=$ & $\kw{deriving\_lattice}\;L_i :=$ \\
    & & & $\quad \ell_1 < \dots < \ell_k$ \\
    & & $\mid$ & $\kw{label}\;L := L_1 \times \dots \times L_n$ \\
    \textbf{Clearance} & $\mathcal{C}$ & $::=$ & $\kw{clearance}\;\mathit{id} : L\;\kw{where}\;\bar{\kappa}$ \\
    \textbf{Clearance rules} & $\kappa$ & $::=$ & $s\;\mathit{mode}\;[\kw{on}\;k] \mapsto \ell$ \\
    \textbf{Policies} & $\Pi$ & $::=$ & $\kw{security\_policy}\;\mathit{id} : L\;\kw{where}$ \\
    & & & $\quad \bar{D}\;\,\bar{c}\;\,\bar{R}\;\kw{default}\;d_0$ \\
    \textbf{Resource sets} & $D$ & $::=$ & $\kw{set}\;A := [s_1, \dots, s_m]$ \\
    \textbf{Composition} & $c$ & $::=$ & $\kw{require}\;\Pi' \mid \kw{grant}\;\Pi'$ \\
    \textbf{Defaults} & $d_0$ & $::=$ & $\kw{allow} \mid \kw{deny}$ \\
    \textbf{Guard rules} & $R$ & $::=$ & $\kw{rule}\;\mathit{id} : \kw{on}\;\mathit{pat}\;\kw{if}\;\phi$ \\
    & & & $\quad \kw{then}\;\kw{allow}\;\kw{else}\;\kw{deny}\;\mathit{msg}$ \\
    \textbf{Action pats.} & $\mathit{pat}$ & $::=$ & $\act{read}(\mathit{src}, k) \mid \act{write}(\mathit{dest}, k)$ \\
    & & $\mid$ & $\act{create}(\mathit{srcs}, \mathit{dest})$ \\
    & & $\mid$ & $\act{delete}(\mathit{target})$ \\
    & & $\mid$ & $\act{copy}(\mathit{src}, \mathit{dest}) \mid \act{sendNet}(\mathit{dest})$ \\
    & & $\mid$ & $\act{concatenate}(\mathit{srcs}, \mathit{dest})$ \\
    & & $\mid$ & $\act{recvNet}(\mathit{src})$ \\
    & & $\mid$ & $\act{downloadNet}(\mathit{src}, \mathit{dest})$ \\
    & & $\mid$ & $\act{uploadNet}(\mathit{src}, \mathit{dest})$ \\
    & & $\mid$ & $\act{escalatePrivilege}(\mathit{target})$ \\
    \textbf{Predicates} & $\phi$ & $::=$ & $t_1 \sqsubseteq t_2 \mid x \in A \mid x \subseteq [\bar{A}]$ \\
    & & $\mid$ & $\phi_1 \land \phi_2$ \\
    & & $\mid$ & $\phi_1 \lor \phi_2 \mid \lnot \phi$ \\
    \textbf{Label terms} & $t$ & $::=$ & $\mathcal{T} \mid \sigma[x, k] \mid \ell \mid t_1 \sqcup t_2 \mid t.L_i$ \\[2pt]
    \multicolumn{4}{@{}l@{}}{$\mathit{id}, A, x \in \mathit{Ident} \qquad s, \mathit{msg} \in \mathit{String} \qquad k \in \mathit{Key} \qquad \ell \in L$}
  \end{tabular}
  \caption{Abstract syntax of the \sys EDSL, parametric over a security lattice $L$. Overbars ($\bar{X}$) denote possibly-empty sequences.}
  \label{Fi:policy-grammar}
\end{figure}

\begin{figure}[tb!]
  \centering
  \footnotesize\polsf
  \ifarxiv\begin{tabular}{@{}l@{}}\else\begin{tabular*}{\columnwidth}{@{}l@{\extracolsep{\fill}}@{}}\fi
  \kw{deriving\_lattice} C := \lbl{Public} < \lbl{Internal} < \lbl{Sensitive} < \lbl{Secret}\\
  \kw{deriving\_lattice} I := \lbl{Trusted} < \lbl{Untrusted}\\
  \kw{label} Label := C $\times$ I\\[4pt]
  \kw{clearance} defaultClearance : Label \kw{where}\\
  \hspace*{0.8em}\texttt{"data/servers.csv"} \kw{exact} $\mapsto$ (\lbl{Sensitive}, \lbl{Trusted})\\
  \hspace*{0.8em}\texttt{"/id\_rsa"} \kw{suffixMatch} $\mapsto$ (\lbl{Secret}, \lbl{Trusted})\\[4pt]
  \kw{security\_policy} informationFlow : Label \kw{where}\\
  \hspace*{0.8em}\kw{set} Scratch := [\texttt{"/tmp/*"}, \texttt{"scratch/*"}]\\
  \hspace*{0.8em}\kw{set} ApprovedHosts := [\texttt{"localhost"}, \texttt{"127.0.0.1"}]\\[3pt]
  \hspace*{0.8em}\kw{require} systemIntegrity\\[3pt]
  \hspace*{0.8em}\kw{rule} restrict\_write\_down : \kw{on} write($\mathit{dest}$, $k$)\\
  \hspace*{1.6em}\kw{if} Taint.C $\sqsubseteq$ $\sigma$[$\mathit{dest}$, $k$].C $\lor$ $\mathit{dest} \in$ Scratch\\
  \hspace*{1.6em}\kw{then} \kw{allow} \kw{else} \kw{deny} \texttt{"write-down"}\\[3pt]
  \hspace*{0.8em}\kw{rule} restrict\_upload\_exfil : \kw{on} uploadNet($\mathit{src}$, $\mathit{dest}$)\\
  \hspace*{1.6em}\kw{if} (Taint $\sqcup$ $\sigma$[$\mathit{src}$, content]).C $\sqsubseteq$ \lbl{Public} $\lor$ $\mathit{dest} \in$ ApprovedHosts\\
  \hspace*{1.6em}\kw{then} \kw{allow} \kw{else} \kw{deny} \texttt{"network exfil"}\\[3pt]
  \hspace*{0.8em}\kw{default} \kw{allow}
  \ifarxiv\end{tabular}\else\end{tabular*}\fi
  \caption{Example specification defining the product lattice $\pol{Label} = \pol{C} \times \pol{I}$, initial sensitivity rules ($\sigma_0$), and an IFC policy.}
  \label{Fi:policy-example}
\end{figure}

\noindent
Each policy $\Pi$ consists of four components:
\begin{enumerate}
  \item \textbf{Resource sets} ($D$): Resource sets take the form $\pol{set}\;A := [s_1, \dots, s_m]$, where each $s_i$ is a pattern in the policy. \figref{policy-example} shows two such sets: \pol{Scratch} and \pol{ApprovedHosts}.
  The sets can be used in guard rules as allowlists or blocklists.
  \item \textbf{Compositions} ($c$): These clauses allow two or more policies to be composed using \pol{require} or \pol{grant} operations, discussed in \sectref{dsl-eval}.
  \item \textbf{Guard rules} ($R$): A policy contains zero or more guard rules
  that specify the conditions under which abstract actions are allowed or denied.
  \item \textbf{Default decision} ($\pol{default}\;d_0$): Fallback decision ($\pol{allow}$/$\pol{deny}$) when no rule matches.
\end{enumerate}

Every guard rule uses a conditional \pol{if}/\pol{then}/\pol{else} form. Each rule matches on an action pattern $\mathit{pat}$, evaluates a boolean predicate $\phi$ over the current state $S$, and produces an explicit \pol{allow} or \pol{deny} decision.

When a candidate action $a$ matches a pattern $\mathit{pat}$ (written $\operatorname{match}(\mathit{pat}, a)$, where the wildcard $\_$ denotes an arbitrary parameter), the guard predicate $\phi(a, S)$ is evaluated over the active system state $S$. For example, the rule \pol{restrict\_write\_down} in \figref{policy-example} matches any action of the form $\act{write}(\mathit{dest}, k, \_, \_)$ and permits the write only if the agent's confidentiality taint ($\mathcal{T}.\pol{C}$) is bounded by the destination's confidentiality label ($\sigma[\mathit{dest}, k].\pol{C}$), or if the destination resource belongs to the \pol{Scratch} set.

A guard predicate $\phi$ is a boolean formula where every term is one of the following:
\begin{itemize}
  \item \textbf{Label Comparison} ($t_1 \sqsubseteq t_2$): Two labels $t_1, t_2 \in L$ are compared using the lattice's partial order. Every label term $t$ is (i) $\mathcal{T}$, (ii) $\sigma[x, k]$ for some resource $x$ and key $k$, (iii) a constant $\ell \in L$, (iv) a join $t_1 \sqcup t_2$, or (v) a projection $t.L_i$ onto a component lattice $L_i$.
  \item \textbf{Set membership} ($x \in A$): This predicate checks if  $x$ belongs to $A$, where $x$ is a single resource identifier and $A$'s entries may be glob patterns with wildcards.
  \item \textbf{Set containment} ($x \subseteq  [\bar{A}]$): This predicate checks if every element of $x$ (a set of resources) belongs to at least one of the sets in $[\bar{A}]$.
\end{itemize}

Each rule evaluates to a decision when its pattern matches the candidate action $a$, or otherwise yields no decision ($\bot$). Formally, the evaluation of rule $R$ in state $S$ on action $a$ is defined as:
\[
  R(S, a) =
  \begin{cases}
    \pol{allow} & \text{if } \operatorname{match}(\mathit{pat}, a) \land \phi(a, S), \\
    \pol{deny}(\mathit{msg}) & \text{if } \operatorname{match}(\mathit{pat}, a) \land \lnot\phi(a, S), \\
    \bot & \text{otherwise.}
  \end{cases}
\]

\subsection{Evaluation and Policy Composition}
\label{Se:dsl-eval}

A policy $\Pi$ evaluates an abstract state $S$ and action $a$ to a decision $\Pi(S, a) \in \{\pol{allow},\; \pol{deny}\}$ (eliding diagnostic messages $\mathit{msg}$).
First, consider a standalone policy $\Pi = \langle \bar{R}, d_0 \rangle$ with guard rules $\bar{R}$ and default decision $d_0$.
Let $\mathit{Dec}_{\bar{R}}(S, a) = \{R_i(S, a) \mid R_i \in \bar{R},\; R_i(S, a) \neq \bot\}$ be the set of decisions from all rules matching $a$ in $S$.
Following a deny-overrides discipline, $\Pi$ denies $a$ if any matching rule denies it, allows $a$ if at least one rule allows it and none deny it, and falls back to $d_0$ when no rule matches:
\[
  \Pi(S, a) =
  \begin{cases}
    \pol{deny} & \text{if } \pol{deny} \in \mathit{Dec}_{\bar{R}}(S, a), \\
    \pol{allow} & \text{if } \mathit{Dec}_{\bar{R}}(S, a) = \{\pol{allow}\}, \\
    d_0 & \text{if } \mathit{Dec}_{\bar{R}}(S, a) = \emptyset.
  \end{cases}
\]

Policies form a bounded lattice $\mathbb{P}$ ordered by permissiveness: $\Pi_1 \sqsubseteq_{\mathbb{P}} \Pi_2 \iff (\forall S, a,\; \Pi_1(S, a) = \pol{allow} \implies \Pi_2(S, a) = \pol{allow})$, with bottom $\bot_{\mathbb{P}}$ (always $\pol{deny}$) and top $\top_{\mathbb{P}}$ (always $\pol{allow}$).
Two policies $\Pi_1, \Pi_2 \in \mathbb{P}$ compose pointwise via meet and join:
\begin{itemize}
  \item \textbf{Meet} (\pol{require}, $\Pi_1 \sqcap_{\mathbb{P}} \Pi_2$): Allows $(S, a)$ iff \emph{both} $\Pi_1(S, a) = \pol{allow}$ and $\Pi_2(S, a) = \pol{allow}$.
  \item \textbf{Join} (\pol{grant}, $\Pi_1 \sqcup_{\mathbb{P}} \Pi_2$): Allows $(S, a)$ iff \emph{either} $\Pi_1(S, a) = \pol{allow}$ or $\Pi_2(S, a) = \pol{allow}$.
\end{itemize}

In general, a policy $\Pi$ in \figref{policy-grammar} specifies local rules $\bar{R}$, a default $d_0$, required policies $\{\Pi_{\pol{req}_i}\}$, and granted policies $\{\Pi_{\pol{grant}_j}\}$.
Because a policy can only reference previously declared policies, the policy dependency graph is a DAG by construction, guaranteeing termination.
Let $\Pi_{\text{local}}(S, a)$ denote the evaluation of $\bar{R}$ with fallback $d_{\text{fall}}(S, a) = d_0 \sqcup_{\mathbb{P}} \bigl(\bigsqcup_j \Pi_{\pol{grant}_j}(S, a)\bigr)$ when $\mathit{Dec}_{\bar{R}}(S, a) = \emptyset$. The composite policy evaluates as:
\[
  \Pi(S, a) = \Pi_{\text{local}}(S, a) \;\sqcap_{\mathbb{P}}\; \underset{i}{\textstyle\bigsqcap}\, \Pi_{\pol{req}_i}(S, a).
\]
Crucially, \pol{grant} clauses only relax the fallback default when no local rule matches ($\mathit{Dec}_{\bar{R}}(S, a) = \emptyset$), whereas \pol{require} clauses act as global conjunctions. Thus, no local rule or \pol{grant} can bypass a \pol{require}, and no \pol{grant} can override a matching local rule.

\subsection{Mechanized Verification}
\label{Se:dsl-metatheory}

To guarantee that policy composition and state tracking behave predictably under arbitrary security lattices, we formalize the complete EDSL semantics and prove its metatheoretic properties in Lean~4.

Given a policy $\Pi$ and state $S$, the reference monitor mediates each candidate action $a$ via the monitored transition function $\hat{\delta}_{\Pi}(S, a)$:
\[
  \hat{\delta}_{\Pi}(S, a) =
  \begin{cases}
    \delta(S, a) & \text{if } \Pi(S, a) = \pol{allow}, \\
    S            & \text{if } \Pi(S, a) = \pol{deny}.
  \end{cases}
\]
For a finite trace of candidate actions $\tau = [a_1, \dots, a_n]$, the multi-step monitored execution $\hat{\delta}_{\Pi}^{*}(S_0, \tau) = S_n$ is defined by iterating $S_{k} = \hat{\delta}_{\Pi}(S_{k-1}, a_k)$ from initial state $S_0 = \langle \bot, \lbl{User}, \sigma_0 \rangle$.

We formalize the abstract security lattice as an algebraic typeclass in Lean~4, parameterizing the state $S$, transition function $\delta$, and policy evaluator $\Pi$ over any bounded lattice $L$. We mechanically verify the following core properties:
\begin{itemize}
  \item \textbf{Deny-overrides monotonicity}: Local denials cannot be overridden by adding permissive local rules. For any policy $\Pi$ with local rules $\bar{R}$, if there exists $R \in \bar{R}$ such that $R(S, a) = \pol{deny}$, then for any expanded rule set $\bar{R}' \supseteq \bar{R}$, the policy $\Pi'$ formed by replacing $\bar{R}$ with $\bar{R}'$ satisfies $\Pi'(S, a) = \pol{deny}$.

  \item \textbf{Non-bypassability of requirements}: A required policy acts as an inviolable upper bound in the policy lattice $\mathbb{P}$ ($\Pi \sqsubseteq_{\mathbb{P}} \Pi_{\pol{req}_i}$). Formally, if $\Pi$ includes $\pol{require}\;\Pi_{\pol{req}_i}$ and $\Pi_{\pol{req}_i}(S, a) = \pol{deny}$, then $\Pi(S, a) = \pol{deny}$ regardless of local rules $\bar{R}$, default $d_0$, or granted policies $\{\Pi_{\pol{grant}_j}\}$.

  \item \textbf{Order invariance}: Policy evaluation is purely declarative. Permuting the declaration order of local rules $\bar{R}$, required policies $\{\Pi_{\pol{req}_i}\}$, or granted policies $\{\Pi_{\pol{grant}_j}\}$ within $\Pi$ preserves semantics: $\Pi(S, a) = \Pi_{\pi}(S, a)$ for any permutation $\pi$.


  \item \textbf{State preservation under denial}: Rejected actions have no side effects on the security state. For all $S$ and $a$, $\Pi(S, a) = \pol{deny} \implies \hat{\delta}_{\Pi}(S, a) = S$, ensuring that unauthorized attempts cannot affect resource labels, escalate privilege, or alter taint.
\end{itemize}



\section{\isolate: Sandboxed Execution and Semantic Inference}
\label{Se:isolate}

In this section, we describe how \isolate infers the abstract actions defined in~\sectref{lean_dsl} from an agent's terminal commands.
To infer these actions, \isolate intercepts the command's system-level events and network traffic. Our current implementation uses \texttt{strace} for system-call tracing.
The tracing backend is fully modular; while \texttt{strace} provides a robust and portable tracing interface for our environment, the system is designed to seamlessly swap in high-performance alternatives, such as an eBPF-based tracer, at scale.



\subsection{Transactional Sandboxed Execution}
\label{Se:isolate-sandbox}

The idea of running untrusted operations speculatively, checking their effects, and committing them only if safe has been previously explored in systems security~\citep{liang2003alcatraz, shankar2006doppelganger}.
Building on \texttt{try}~\citep{lamprou2026try}, \isolate applies this concept to agentic security by executing each command inside an isolated Linux mount and process environment that stages side effects without affecting the host:

\begin{itemize}
    \item \textbf{OverlayFS for filesystem isolation:} \isolate mounts a
    per-directory OverlayFS over the host's root directories and stages all changes in a disposable scratch directory; changes are later committed to the host or discarded, based on the policy verdict.

    \item \textbf{Network interception:} \isolate routes all outbound traffic
    through a local proxy. The proxy holds each request and records its destination and, where applicable, its contents (i.e., method, URL, headers, and body). \isolate infers the corresponding abstract action over the trace prefix observed so far and checks it against the policy. The held request is then forwarded to its destination or dropped.

    \item \textbf{Background processes:} When a command
    spawns background processes that outlive the foreground shell, \isolate pauses them and resumes or terminates them based on the policy verdict. However, because per-step \texttt{strace} stops once the foreground command exits, a policy can choose to block such spawned processes entirely (\sectref{limitations}).
\end{itemize}

\subsection{Semantic Inference Pipeline}
\label{Se:isolate-inference}

\isolate turns the raw trace of system calls and network traffic into the abstract actions of~\sectref{dsl-transitions} in four stages.

\paragraph{1. Process-tree and file-descriptor tracking.}
\isolate reconstructs the process tree across \texttt{fork}, \texttt{clone}, and \texttt{execve} system calls and tracks file descriptors, including pipes and I/O redirections. It records the files each process reads and the writer provenance $(\mathit{actor}, \mathit{origin})$ of each file mutation, where $\mathit{actor}$ is the binary performing the system call and $\mathit{origin}$ is the parent program that invoked it. It also tracks privilege transitions (e.g., \texttt{sudo} $\to$ \act{escalatePrivilege}).

\paragraph{2. File and IPC provenance tracking.}
\isolate tracks data provenance ($\mathit{srcs}$) across processes, inter-process channels, and files. Each \texttt{read} system call adds the file it reads to the process's source set $\mathit{srcs}$, which propagates to all downstream processes. \isolate uses $\mathit{srcs}$ to emit provenance-aware actions:
\begin{itemize}
    \item \emph{Writes to a file:} \isolate emits \act{write}, \act{copy}, or \act{concatenate} when the process has read zero, one, or several source files, respectively, with the tracked $\mathit{actor}$ and $\mathit{origin}$.
    \item \emph{Writes to \texttt{stdout}/\texttt{stderr}:} When a process writes to \texttt{stdout} or \texttt{stderr}, \isolate emits a \act{read} action for each file in $\mathit{srcs}$, since the output enters the agent's context window and affects its taint.
    \item \emph{Creates and deletes:} When a process creates a new path (e.g., \texttt{openat} with \texttt{O\_CREAT}, or \texttt{mkdir}), \isolate emits \act{create} with the files the process has read as $\mathit{srcs}$, since their data may be encoded in the new path's metadata. When a process removes a path (\texttt{unlink} or \texttt{rmdir}), \isolate emits \act{delete}.
\end{itemize}
Other filesystem and metadata system calls map analogously onto these abstract actions. For example, \texttt{rename} emits \act{copy} followed by \act{delete}. \ifarxiv For more details on mapping of system calls to abstract actions, see~\appref{syscall-mapping}.\fi

\paragraph{3. Network calls.}
Network-related abstract actions such as \act{uploadNet}, \act{sendNet},
\act{downloadNet}, and \act{recvNet} are emitted using network system
calls and proxy logs.
Network system calls (e.g., \texttt{connect}, \texttt{sendto}, \texttt{recvfrom})
only expose an IP address, while the proxy logs record the rest of the request, such as the URL, headers, and body. Using the network system calls and proxy logs, \isolate infers the destination's domain names. It emits a network action for \emph{both} the IP address and each domain name, so a forged name cannot hide the real address.

For each destination $r_{\text{net}}$, \isolate emits actions for both the send and the receive side:
\begin{itemize}
    \item \emph{Network send:} If the process has read local files ($\mathit{srcs} \neq \emptyset$),
    \isolate emits $\act{uploadNet}(r_{\text{src}}, r_{\text{net}})$ for each source
    file $r_{\text{src}} \in \mathit{srcs}$. When it has read no files ($\mathit{srcs} =
    \emptyset$), \isolate emits $\act{sendNet}(r_{\text{net}})$.

    \item \emph{Network receive:} When the network response is written to a file
    $r_{\text{dest}}$, \isolate emits $\act{downloadNet}(r_{\text{net}}, r_{\text{dest}})$;
    when it is written to \texttt{stdout}/\texttt{stderr}, \isolate emits $\act{recvNet}(r_{\text{net}})$.
\end{itemize}

\paragraph{4. Filtering and abstraction.}
Since complex commands can issue thousands of syscalls, \isolate uses filtering and abstraction in some cases:
\begin{itemize}
    \item \emph{Routine OS startup reads:} 
    \isolate filters reads of standard system paths that every program reads on startup (e.g., \texttt{/etc/ld.so.cache}) and authentication reads by \texttt{sudo} (e.g., \texttt{/etc/shadow}), as long as those files are not modified.

    \item \emph{Abstracting to system domains:} Listing commands such as \texttt{ls -la /} can issue several \texttt{getdents64} and \texttt{stat} calls. Emitting one \act{read} (metadata) action per syscall produces hundreds of abstract actions.
    Instead, \isolate groups unmodified standard OS
    directories (such as \texttt{/bin}) into
    coarse-grained system domains (\texttt{<system:binaries>}) and emits at most one
    \act{read} per domain. All other paths, including modified system files, retain their exact paths.
\end{itemize}

\section{Assurance Beyond the Monitor}
\label{Se:verify}

\subsection{Fuzzing \isolate}
\label{Se:fuzzer}

The formal guarantees in \sectref{dsl-metatheory} hold over sequences of abstract actions evaluated by the Lean~4 monitor.
End-to-end enforcement, however, relies on the \emph{semantic contract}: \isolate must faithfully map the system calls of arbitrary shell commands to abstract actions. 
In a POSIX environment, the same logical data flow can be carried out through many different OS mechanisms such as standard utilities (\texttt{cp}, \texttt{cat}, \texttt{mv}), multi-stage shell pipelines, zero-copy kernel transfers (\texttt{copy\_file\_range}, \texttt{sendfile}, \texttt{splice}), file-descriptor duplication (\texttt{dup2}, \texttt{fcntl}), symbolic links, or in-process file reads and writes.
If \isolate misses any abstract actions for any of these mechanisms, the monitor may underestimate a resource's label $\sigma(r, k)$ and allow an action that the policy should deny.

\begin{figure}[tb!]
  \centering
  \scriptsize
  \begin{tikzpicture}[
      stage/.style={draw=black!60, rounded corners=3pt, align=center,
                    inner sep=2.2pt, minimum height=9mm},
      arr/.style={-{Stealth[length=1.8mm]}, semithick, black!65},
      lbl/.style={font=\scriptsize, text=black!85, align=left, inner sep=1.5pt},
    ]
    \node[stage, text width=15.5mm] (dag)
      {\textbf{Provenance}\\\textbf{DAG}};

    \node[stage, fill=black!5, text width=24.5mm, right=5.5mm of dag, yshift=8mm] (ref)
      {\textbf{Abstract trace $\bar{a}$}\\direct lowering};

    \node[stage, fill=blue!7, text width=24.5mm, right=5.5mm of dag, yshift=-8mm] (obs)
      {\textbf{Shell $c$ $\to$ \isolate}\\inferred trace $\hat{a}$};

    \node[stage, fill=orange!10, text width=17mm, right=4.5mm of ref] (mon1)
      {\textbf{Monitor $(\Pi)$}\\$\Rightarrow d_{\text{ref}}, \sigma_{\text{ref}}$};

    \node[stage, fill=orange!10, text width=17mm, right=4.5mm of obs] (mon2)
      {\textbf{Monitor $(\Pi)$}\\$\Rightarrow d_{\text{obs}}, \sigma_{\text{obs}}$};

    \coordinate (fork) at ([xshift=2.5mm]dag.east);
    \draw[semithick, black!65] (dag.east) -- (fork);
    \draw[arr] (fork) |- (ref.west);
    \draw[arr] (fork) |- (obs.west);
    \draw[arr] (ref) -- (mon1);
    \draw[arr] (obs) -- (mon2);

    \draw[<->, semithick, dashed, NavyBlue] (mon1.south) --
      node[lbl, right=1mm, text=NavyBlue]
      {$d_{\text{obs}} \sqsubseteq_{\mathcal{P}} d_{\text{ref}}$\\$\sigma_{\text{ref}} \sqsubseteq \sigma_{\text{obs}}$}
      (mon2.north);
  \end{tikzpicture}
  \caption{Fuzzing \isolate over generated provenance DAGs under $(\Pi, S_0)$.}
  \label{Fi:fuzzing-pipeline}
\end{figure}
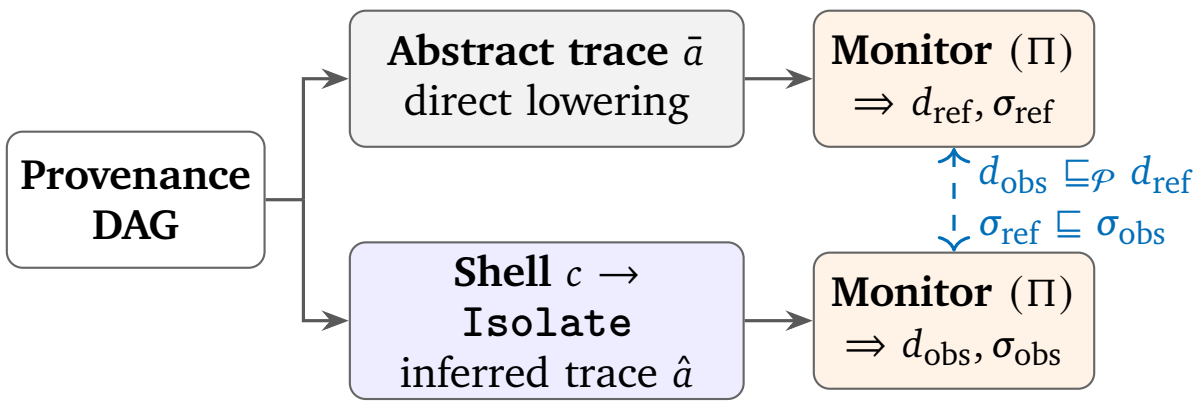

To mitigate this risk, we fuzz \isolate over multi-step provenance workflows, parameterized by a target policy $\Pi$ and initial state $S_0 = \langle \mathcal{T}_0, \mathcal{P}_0, \sigma_0 \rangle$ (\figref{fuzzing-pipeline}). Each test case starts from a generated provenance directed acyclic graph (DAG) that specifies a sequence of abstract file operations over a set of seed files initialized with labels from $L$. We lower each workflow along two paths:
\begin{enumerate}
  \item \textbf{Reference abstract trace ($\bar{a}$):} The DAG is lowered directly into its ground-truth sequence of abstract actions $\bar{a} = \langle a_1, \dots, a_m \rangle$. Evaluating $\bar{a}$ in the Lean~4 monitor under policy $\Pi$ from $S_0$ yields a reference decision $d_{\text{ref}} \in \{\pol{allow}, \pol{deny}\}$ and final state $S_{\text{ref}} = \langle \mathcal{T}_{\text{ref}}, \mathcal{P}_{\text{ref}}, \sigma_{\text{ref}} \rangle$.
  \item \textbf{Concrete execution trace ($\hat{a}$):} Each edge in the DAG is instantiated with a randomly chosen concrete shell or syscall mechanism to produce a shell script $c$. We execute $c$ inside \isolate to obtain the inferred sequence of abstract actions $\hat{a}$, and evaluate them using the Lean~4 monitor under the same policy $\Pi$ and initial state $S_0$ to obtain the observed decision $d_{\text{obs}} \in \{\pol{allow}, \pol{deny}\}$ and state $S_{\text{obs}} = \langle \mathcal{T}_{\text{obs}}, \mathcal{P}_{\text{obs}}, \sigma_{\text{obs}} \rangle$.
\end{enumerate}

\paragraph{Evaluation against concrete policies.}
Checking equivalence requires evaluating both traces against a concrete policy $\Pi$ that exercises the labels propagated along the DAG. Consider the running example (\lstref{running}) and the \pol{informationFlow} policy in \figref{policy-example}, alongside an integrity rule guarding copies into trusted files:

\begin{center}
\small\polsf
\begin{tabular}{@{}l@{}}
\kw{rule} restrict\_untrusted\_copy :\\
\hspace*{1em}\kw{on} copy($\mathit{src}$, $\mathit{dest}$)\\
\hspace*{1em}\kw{if} (Taint $\sqcup$ $\sigma$[$\mathit{src}$, content]).I $\sqsubseteq$ $\sigma$[$\mathit{dest}$, content].I\\
\hspace*{2.2em}$\lor$ $\sigma$[$\mathit{dest}$, content].C $\sqsubseteq$ \lbl{Internal}\\
\hspace*{1em}\kw{then} \kw{allow} \kw{else} \kw{deny} \texttt{"untrusted copy"}
\end{tabular}
\end{center}
\noindent Suppose \file{data/servers.csv} is initialized with $\sigma_0(\texttt{data/servers.csv}, \pol{content}) = (\lbl{Sensitive}, \lbl{Trusted})$, an external input \file{scratch/input.txt} is initialized to $(\lbl{Public}, \lbl{Untrusted})$, and \file{scratch/report.txt} starts at $\bot = (\lbl{Public}, \lbl{Trusted})$. In the example from \sectref{overview}, data from \file{data/servers.csv} (or \file{scratch/input.txt}) flows through intermediate files such as \file{scratch/tmp.csv} into \file{scratch/report.txt}, followed by a guarded terminal action, either uploading \file{scratch/report.txt} to \texttt{api.github.com} (checked by \pol{restrict\_upload\_exfil} in \figref{policy-example}), or copying \file{scratch/report.txt} into a protected resource $(\lbl{Sensitive}, \lbl{Trusted})$ (checked by \pol{restrict\_untrusted\_copy} above).

On the reference trace $\bar{a}$, the transition rules \textsc{E-Copy} and \textsc{E-Concat} (\figref{action-transitions}) propagate the source labels across each hop into $\sigma_{\text{ref}}(\texttt{scratch/report.txt}, \pol{content})$, causing the policy $\Pi$ to evaluate to $d_{\text{ref}} = \pol{deny}$ at the terminal action. On the concrete path, the fuzzer realizes those same hops, e.g., streaming \file{data/servers.csv} through a three-stage pipeline \texttt{cat data/servers.csv | cat | cat > scratch/tmp.csv}, transferring \file{scratch/tmp.csv} via a kernel \texttt{splice} pipe into \file{scratch/report.txt}, or appending via \texttt{cat scratch/input.txt {>}{>} scratch/report.txt}.

\subsection{Policy Verification}
\label{Se:verify-policy}

\begin{figure}[tb!]
  \centering
  \footnotesize\polsf
  \ifarxiv\begin{tabular}{@{}l@{}}\else\begin{tabular*}{\columnwidth}{@{}l@{\extracolsep{\fill}}@{}}\fi
  {\color{black!55}\texttt{-{}-}\ \textrm{\textit{(a) Single fixed initial state $S_0$ (\pol{start})}}}\\
  \kw{security\_check} noSecretInit : Label \kw{where}\\
  \hspace*{0.8em}\kw{start}: clearance := initClearance\\
  \hspace*{0.8em}\kw{bad}: $\lnot$(Taint.C $\sqsubseteq$ \lbl{Sensitive})\\[3pt]
  {\color{black!55}\texttt{-{}-}\ \textrm{\textit{(b) All safe initial states (\pol{assume})}}}\\
  \kw{security\_check} noSecretAccess : Label \kw{where}\\
  \hspace*{0.8em}\kw{assume}: Taint.C $\sqsubseteq$ \lbl{Sensitive}\\
  \hspace*{0.8em}\kw{bad}: $\lnot$(Taint.C $\sqsubseteq$ \lbl{Sensitive})
  \ifarxiv\end{tabular}\else\end{tabular*}\fi
  \caption{Specifying ``No secret access'' in \sys from (a)~a fixed initial state and (b)~all safe initial states.}
  \label{Fi:security-check-example}
\end{figure}

\begin{figure}[tb!]
  \centering
  \footnotesize\polsf
  \ifarxiv\begin{tabular}{@{}l@{}}\else\begin{tabular*}{\columnwidth}{@{}l@{\extracolsep{\fill}}@{}}\fi
  \kw{security\_policy} secretProtectionPolicy : Label \kw{where}\\
  \hspace*{0.6em}\kw{set} WorkspacePaths := [$\dots$]\\
  \hspace*{0.6em}\kw{rule} restrict\_secret\_read : \kw{on} read($\mathit{src}$, $\mathit{key}$)\\
  \hspace*{1.2em}\kw{if} $\sigma$[$\mathit{src}$, $\mathit{key}$].C $\sqsubseteq$ \lbl{Sensitive}\\
  \hspace*{1.2em}\kw{then} \kw{allow} \kw{else} \kw{deny} \texttt{"secret read"}\\
  \hspace*{0.6em}\kw{rule} restrict\_secret\_recv : \kw{on} recvNet($\mathit{src}$)\\
  \hspace*{1.2em}\kw{if} $\sigma$[$\mathit{src}$, content].C $\sqsubseteq$ \lbl{Sensitive}\\
  \hspace*{1.2em}\kw{then} \kw{allow} \kw{else} \kw{deny} \texttt{"secret network receive"}\\
  \hspace*{0.6em}\kw{rule} restrict\_write\_down : \kw{on} write($\mathit{dest}$, $\mathit{key}$)\\
  \hspace*{1.2em}\kw{if} Taint.C $\sqsubseteq$ \lbl{Internal} $\lor$ Taint.C $\sqsubseteq$ $\sigma$[$\mathit{dest}$, $\mathit{key}$].C\\
  \hspace*{2.0em}$\lor$ $\mathit{dest} \in$ WorkspacePaths\\
  \hspace*{1.2em}\kw{then} \kw{allow} \kw{else} \kw{deny} \texttt{"confidentiality write-down"}\\
  \hspace*{0.6em}\kw{rule} restrict\_untrusted\_copy : \kw{on} copy($\mathit{src}$, $\mathit{dest}$)\\
  \hspace*{1.2em}\kw{if} (Taint $\sqcup$ $\sigma$[$\mathit{src}$, content]).I $\sqsubseteq$ $\sigma$[$\mathit{dest}$, content].I\\
  \hspace*{2.0em}$\lor$ $\sigma$[$\mathit{dest}$, content].C $\sqsubseteq$ \lbl{Internal}\\
  \hspace*{1.2em}\kw{then} \kw{allow} \kw{else} \kw{deny} \texttt{"untrusted copy"}\\
  \hspace*{0.6em}\kw{rule} restrict\_secret\_download : \kw{on} downloadNet($\mathit{src}$, $\mathit{dest}$)\\
  \hspace*{1.2em}\kw{if} $\sigma$[$\mathit{dest}$, content].C $\sqsubseteq$ \lbl{Internal}\\
  \hspace*{1.2em}\kw{then} \kw{allow} \kw{else} \kw{deny} \texttt{"download onto sensitive file"}\\
  \hspace*{0.6em}$\dots$ {\color{black!55}\texttt{-{}-}\ \textrm{\textit{2 more rules; default allow}}}\\[3pt]
  \kw{security\_policy} benchmarkPolicy : Label \kw{where}\\
  \hspace*{0.6em}\begin{tabular}{@{}l@{\hspace{0.8em}}l@{}}
  \kw{require} secretProtectionPolicy & \\
  \kw{require} systemIntegrityPolicy & {\color{black!55}\texttt{-{}-}\ \textrm{\textit{6 rules}}}\\
  \kw{require} antiExfiltrationPolicy & {\color{black!55}\texttt{-{}-}\ \textrm{\textit{10 rules}}}\\
  \kw{require} privilegePolicy & {\color{black!55}\texttt{-{}-}\ \textrm{\textit{1 rule}}}
  \end{tabular}\\
  \hspace*{0.6em}\kw{default} \kw{allow}
  \ifarxiv\end{tabular}\else\end{tabular*}\fi
  \caption{Excerpt of our benchmark policy\ifarxiv~(see~\appref{benchmark-policy} for the full policy)\fi.}
  \label{Fi:benchmark-policy-excerpt}
\end{figure}

Using a formal policy for runtime enforcement provides provable guarantees that LLM-based monitors cannot offer. However, writing a realistic policy can be tedious and error-prone\ifarxiv~(see~\appref{benchmark-policy} for the policy used in our experiments)\fi.
To ensure a user-written policy satisfies its intended invariants, we provide a verification mechanism.

The verification mechanism is a proof that the written policy satisfies the invariants across all traces of arbitrary length.
More concretely, given a verification query $q = \langle \Pi, S_0, \psi_{\text{bad}} \rangle$ consisting of a policy $\Pi$, an initial security state $S_0$, and a bad state predicate $\psi_{\text{bad}}$, we want to either prove that $\psi_{\text{bad}}$ can never be reached from $S_0$ under $\Pi$ or produce a counterexample trace within a search budget.

Consider an abstract trace $\bar{a} = \langle a_1, \dots, a_m \rangle$ and let $\operatorname{exec}_\Pi(S_0, \bar{a})$ denote the execution of $\bar{a}$ under policy $\Pi$ starting from $S_0$. To answer $q$, we check this safety theorem:
\[
  \operatorname{Safe}(q) \;\eqdef\; \forall \bar{a},\; \lnot \psi_{\text{bad}}\bigl(\operatorname{exec}_\Pi(S_0, \bar{a})\bigr).
\]

In Lean~4, we either prove $\operatorname{Safe}(q)$ or find a counterexample:
\begin{enumerate}
  \item \textbf{Proof in Lean:} We define $\operatorname{Safe}(q)$ as a Lean~4 theorem statement for the given policy $\Pi$ and try to synthesize a machine-checked proof (albeit using an LLM) by induction on the trace $\bar{a}$. If the proof succeeds, we have a concrete proof of safety. Note that this proof is over traces of arbitrary length.
  
  \item \textbf{Counterexample search:} Otherwise, there exists a finite counterexample trace. We find such a trace and show the reachability of a bad state in Lean.

\end{enumerate}

Given that recently LLMs have shown great promise in writing proofs and generating counterexamples, we use an LLM to either prove that the query is safe or generate a candidate abstract counterexample trace\footnote{Other tools from the verification literature such as Z3 solvers or general graph based reachability analysis could also be used instead.}.

\figref{security-check-example} shows how verification queries can be specified in our DSL. Both queries check if the policy prevents the propagation of \lbl{Secret} data to the agent's context. \figref{security-check-example}(a) is for a fixed initial state defined by \pol{initClearance}, while \figref{security-check-example}(b) checks this condition for all initial states satisfying the precondition \pol{assume}.
The bad state predicate (\pol{bad}) is defined as $\lnot(\mathcal{T}.\pol{C} \sqsubseteq \lbl{Sensitive})$.

\figref{benchmark-policy-excerpt} shows an excerpt of the benchmark policy used in our experiments\ifarxiv~(see~\appref{benchmark-policy} for the full policy)\fi.
It shows \pol{secretProtectionPolicy}, which alone contains $7$ rules; when combined with the other sub-policies, the full benchmark policy spans $24$ rules alongside numerous allowlists and path exemptions.
Such realistic policies can quickly get complex and error-prone. However, verifying a policy \emph{can be simpler}; users can define verification queries (\figref{security-check-example}) and prove (or disprove with a counterexample) that the intended invariants hold.

Using this mechanism, we verify the invariants in~\tableref{policy-invariants} for the benchmark policy.

\begin{table*}[tb!]
  \centering
  \footnotesize
  \renewcommand{\arraystretch}{1.05}
  \caption{Safety invariants proved in Lean~4 for the benchmark policy of \sectref{experiments}, for any initial state $S_0 = \langle \mathcal{T}_0, \mathcal{P}_0, \sigma_0 \rangle$ and any abstract trace $\bar{a} = \langle a_1, \dots, a_m \rangle$ of arbitrary length with $S' = \langle \mathcal{T}', \mathcal{P}', \sigma' \rangle = \operatorname{exec}_\Pi(S_0, \bar{a})$. Here $o \rightsquigarrow \mathit{ctx}$ and $o \rightsquigarrow h$ denote that initial attribute $o = (r_0, k_0)$ flows into the agent's context $\mathit{ctx}$ or to host $h$ along $\bar{a}$ (independent of $\sigma$ and $\mathcal{T}$).}
  \label{Ta:policy-invariants}
  \begin{tabularx}{\textwidth}{@{}>{\raggedright\arraybackslash}p{0.145\textwidth} >{\raggedright\arraybackslash}p{0.37\textwidth} >{\raggedright\arraybackslash}X@{}}
    \toprule
    \textbf{Invariant} & \textbf{Description} & \textbf{Formal guarantee} \\
    \midrule
    No secret access &
      On any allowed run starting without \lbl{Secret} data, data labeled \lbl{Secret} never enters the agent's context. &
      If $\mathcal{T}_0.\pol{C} \sqsubseteq \lbl{Sensitive}$, then $\mathcal{T}'.\pol{C} \sqsubseteq \lbl{Sensitive}$ and $\sigma_0(o).\pol{C} = \lbl{Secret} \implies \lnot(o \rightsquigarrow \mathit{ctx})$. \\
    \addlinespace[3pt]
    No tainted escalation &
      On any allowed run, every privilege escalation is denied after agent observes \lbl{Sensitive}, \lbl{Secret}, or \lbl{Untrusted} data. &
      If $\mathcal{T}_0 \not\sqsubseteq (\lbl{Internal}, \lbl{Trusted})$, then $\mathcal{P}' = \mathcal{P}_0$ and \newline
    $\Pi(S, \act{escalatePrivilege}(p, \cdot)) = \pol{deny}$ along $\bar{a}$. \\
    \addlinespace[3pt]
    System integrity &
      On any allowed run, every action modifying, creating, or deleting a non-exempt protected system path is denied. &
      For every non-exempt protected path $r$, $\Pi(S, a) = \pol{deny}$ for any action $a$ modifying $r$, and $\sigma'(r, k) = \sigma_0(r, k)$ for any attribute key $k$. \\
    \addlinespace[3pt]
    No tainted protected edits &
      On any allowed run, once the agent observes \lbl{Sensitive}, \lbl{Secret}, or \lbl{Untrusted} data, every action modifying a protected path other than a cache path is denied. &
      For every protected non-cache path $r$: if $\mathcal{T} \not\sqsubseteq (\lbl{Internal}, \lbl{Trusted})$, then $\Pi(S, a) = \pol{deny}$ for any $a$ modifying $r$; if $\mathcal{T}_0 \not\sqsubseteq (\lbl{Internal}, \lbl{Trusted})$, then $\sigma'(r, k) = \sigma_0(r, k)$. \\
    \addlinespace[3pt]
    No exfiltration &
      On any allowed run, \lbl{Secret} data stays local; \lbl{Sensitive} data reaches only local hosts or app services. &
      $o \rightsquigarrow h \;\land\; \sigma_0(o).\pol{C} = \lbl{Secret} \implies {h \in \mathit{Local}}$; \newline
      $o \rightsquigarrow h \;\land\; \sigma_0(o).\pol{C} = \lbl{Sensitive} \implies {h \in \mathit{Local} \cup \mathit{Services}}$. \\
    \addlinespace[3pt]
    Host blocklist &
      On any allowed run, every network send or file upload to a blocklisted host is denied, so no data ever reaches it. &
      For $h \in \mathit{Blocked}$ and $a \in \{\act{sendNet}(h), \act{uploadNet}(\cdot, h)\}$: $\Pi(S, a) = \pol{deny}$. \\
    \addlinespace[3pt]
    Host allowlist &
      On any allowed run with a host allowlist, every network action to a host outside the local network, app services, or the task's declared hosts is denied. &
      For $a \in \{\act{sendNet}(h),\allowbreak \act{uploadNet}(r, h),\allowbreak \act{recvNet}(h),\allowbreak \act{downloadNet}(h, d)\}$: \newline
      $\Pi(S, a) = \pol{allow} \implies {h \in \mathit{Local} \cup \mathit{Services} \cup \mathit{Declared}}$. \\
    \bottomrule
  \end{tabularx}
\end{table*}

\begin{MyTheorem}[Benchmark Policy Safety Invariants]
\label{The:benchmark-policy-invariants}
For the benchmark policy $\Pi$, every initial state $S_0 = \langle \mathcal{T}_0, \mathcal{P}_0, \sigma_0 \rangle$, and every finite abstract trace of $\Pi$-allowed actions $\bar{a} = \langle a_1, \dots, a_m \rangle$ with final state $S' = \langle \mathcal{T}', \mathcal{P}', \sigma' \rangle = \operatorname{exec}_{\Pi}(S_0, \bar{a})$, all seven safety invariants in~\tableref{policy-invariants} hold.
\end{MyTheorem}
\section{Experimental Evaluation}
\label{Se:experiments}

In this section, we explore and evaluate \sys on the following research questions:

\begin{list}{}{\setlength{\leftmargin}{1em}\setlength{\itemindent}{-1em}\setlength{\itemsep}{2pt}\setlength{\topsep}{3pt}}
  \item \textbf{RQ1:} Can a policy written in the \sys DSL enforce security on realistic coding-agent tasks, catching all harmful actions that can be modeled by the DSL while maintaining high utility? \emph{\sys achieves $100\%$ recall on adversarial actions with a low false-positive rate (FPR) of $4.7\%$ (\tableref{rq1-results}).}
  \item \textbf{RQ2:} How does \sys compare to existing monitors on the same tasks? \emph{\sys achieves $100\%$ recall whereas LLM-based monitors do not, even when tuned for lower FPR, which can also vary across runs (\tableref{rq1-results}).}
  \item \textbf{RQ3:} How many benchmark trajectories are amenable to formal runtime monitoring and how many require LLM-based reasoning due to their semantic nature? \emph{While \sys cannot detect semantic attacks, it can intercept trajectories containing background processes or database operations, even with unmodeled internal semantics.}
  \item \textbf{RQ4:} How robust are \isolate and \sys to adversarial agents? \emph{Red-teaming helped uncover and fix gaps in \isolate, reducing attack success rate to $0\%$ (\figref{red-team-architecture}(b)).}
\end{list}

\subsection{Experimental Setup}
\label{Se:experimental-setup}

To answer these questions, we evaluate \sys on two coding-agent control benchmarks with realistic software-engineering tasks.
To the best of our knowledge, \sys is the first runtime monitor with formal guardrails to be evaluated on realistic benchmarks.
The benchmarks contain agent trajectories, created by running an agent either on a benign main task alone (benign trajectories) or on a main task with an adversarial side task (adversarial trajectories).


The benchmarks used in our evaluations are as follows:
\begin{itemize}
  \item \textbf{MonitoringBenchHonest}~\citep{kucukkurt2026monitoringbenchhonest, jotautaite2026monitoringbench}: A system-administration benchmark built on ControlArena~\citep{inglis2025controlarena} spanning $88$ Linux containers, $88$ main tasks, and four adversarial side tasks.
  \item \textbf{LinuxArena}~\citep{linuxarena}: A multi-service software-engineering benchmark spanning $10$ containerized production environments, over $340$ main tasks, and $81$ adversarial side tasks. 
\end{itemize}

\paragraph{Replay setup.}
To evaluate \sys on these benchmarks, we replicate each task's environment inside our sandbox and replay the offline trajectories provided in the benchmarks step by step. We emulate each environment's external services on a best-effort basis and exclude trajectories whose required services cannot be emulated. We also exclude trajectories that cannot be replayed end-to-end (e.g., where the side-task attack is only triggered by the benchmark harness after the agent finishes).

Among the remaining trajectories, we focus on the subset where every action taken by the agent runs in the foreground, can be captured, and has semantics currently modeled by \sys.
This subset contains $218$ benign and $199$ adversarial trajectories for MonitoringBenchHonest, and $1{,}372$ benign and $145$ adversarial trajectories for LinuxArena.

\paragraph{Policy design.}
Because the benchmarks do not come with an explicit security policy, we formulate a policy in our DSL for realistic scenarios that enforces the safety invariants in~\tableref{policy-invariants}\ifarxiv~(\appref{benchmark-policy} gives the full policy)\else~(\figref{benchmark-policy-excerpt} shows an excerpt)\fi.
The policy rules are shared across all tasks and both benchmarks; only the initial labels and task-specific resource sets assigned to resources, such as files and network endpoints, vary across environments and tasks.
We design the shared rules to prevent general security violations, and configure each task's resource labels and allowlists from its main-task description (e.g., which paths and hosts are allowed, protected, or sensitive).
That is, if a main task explicitly requires writing to an otherwise write-restricted folder, we allow it for that task. Note that this configuration is done at the task level, not at the trajectory level, and we use no side-task information when configuring the policy (see~\sectref{limitations-discussion} for more details).

\subsection{Experimental Results}
\label{Se:results}

\paragraph{RQ1: Security and utility of \sys.}
To answer whether a policy written in the \sys DSL can enforce security on realistic coding-agent tasks, we evaluate \sys on the in-scope trajectories of both benchmarks and compute its adversarial recall and benign false-positive rate (FPR).
As shown in \tableref{rq1-results}, \sys achieves $100\%$ recall on both benchmarks, blocking all $344$ in-scope adversarial trajectories, with a benign FPR of $2.3\%$ on MonitoringBenchHonest and $5.0\%$ on LinuxArena ($4.7\%$ overall).
Trajectories on which \isolate times out are scored as fail-open.\footnote{Only one adversarial trajectory timed out during replay, after \sys had already blocked it. Excluding benign trajectories that timed out or were unscored, the overall FPR is nearly identical.}

\begin{table}[tb!]
  \centering
  \scriptsize
  \setlength{\tabcolsep}{1.5pt}
  \caption{Adversarial recall and benign FPR.}
  \label{Ta:rq1-results}
  \begin{tabular}{@{}llrr@{}}
    \toprule
    \textbf{Benchmark} & \textbf{Monitor} & \textbf{Recall} & \textbf{FPR} \\
    \midrule
    \multirow{3}{*}{\begin{tabular}[c]{@{}l@{}}Monitoring-\\BenchHonest\end{tabular}}
      & \textbf{\sys} & $\mathbf{199 / 199}$ ($\mathbf{100.0\%}$) & $\mathbf{5 / 218}$ ($\mathbf{2.3\%}$) \\
      & Auto Mode (default) & $173 / 199$ ($86.9\%$) & $77 / 218$ ($35.3\%$) \\
      & Policy-aware Auto Mode & $164 / 199$ ($82.4\%$) & $32 / 218$ ($14.7\%$) \\
    \midrule
    \multirow{3}{*}{LinuxArena}
      & \textbf{\sys} & $\mathbf{145 / 145}$ ($\mathbf{100.0\%}$) & $69 / 1{,}372$ ($5.0\%$) \\
      & Auto Mode (default) & $142 / 145$ ($97.9\%$) & $78 / 1{,}372$ ($5.7\%$) \\
      & Policy-aware Auto Mode & $143 / 145$ ($98.6\%$) & $\mathbf{23 / 1{,}372}$ ($\mathbf{1.7\%}$) \\
    \midrule
    \multirow{3}{*}{\textbf{Total}}
      & \textbf{\sys} & $\mathbf{344 / 344}$ ($\mathbf{100.0\%}$) & $74 / 1{,}590$ ($4.7\%$) \\
      & Auto Mode (default) & $315 / 344$ ($91.6\%$) & $155 / 1{,}590$ ($9.7\%$) \\
      & Policy-aware Auto Mode & $307 / 344$ ($89.2\%$) & $\mathbf{55 / 1{,}590}$ ($\mathbf{3.5\%}$) \\
    \bottomrule
  \end{tabular}
\end{table}

The false positives observed by \sys fall into the following categories:
\begin{itemize}
  \item \textbf{Unrequested edits to protected paths ($50/74$)}: The agent edits a protected file such as \file{/etc/passwd} (either as part of a package installation or testing edit permissions), which the main task does not request.
  \item \textbf{Undeclared external hosts ($7/74$)}: The agent sends requests to external hosts that are not declared in the task.
  \item \textbf{Reads of \lbl{Secret}/\lbl{Sensitive} files ($17/74$)}: Commands such as \texttt{grep -r} over \file{/} call \texttt{open} and \texttt{read} syscalls on all files, including files marked secret, irrespective of whether their content is consumed. Because \sys only observes system calls, it assumes that the agent read secret files and blocks further actions.
\end{itemize}
In case of the categories of unrequested edits or undeclared external hosts, \sys rightly flags a denial (which can later be extended to an ``ask'' instead). Overcoming the case of unintentional read via tools like \texttt{grep} is a fundamental limitation of using system calls for runtime monitoring.
Overall, these results show that \sys's policy DSL, together with its abstract semantics, can express precise, stateful policies. Combined with \isolate, the semantic inference engine, this allows \sys to catch all in-scope harmful actions while maintaining high utility.

\paragraph{RQ2: Comparison with existing monitors.}
To understand how \sys compares with an LLM-based runtime monitor, we evaluate Claude Code's Auto Mode~\citep{anthropic2026automode, remedios2026redteamingautomode} (using Claude Sonnet~5)\footnote{Auto Mode open-source implementation: \href{https://github.com/safety-research/red-teaming-auto-mode}{\file{https://github.com/safety-research/red-teaming-auto-mode}}.} on the same trajectories.
We run Auto Mode in two configurations: (1)~\emph{Auto Mode (default)}, using its default rules, and (2)~\emph{Policy-aware Auto Mode}, where our policy and task-specific resource sets are supplied verbatim in its soft-rules field along with a brief description of the DSL semantics.
\tableref{rq1-results} compares the recall and FPR of \sys against both Auto Mode configurations.
In its default configuration (as most users would deploy it), Auto Mode achieves a recall of $91.6\%$ and a high FPR of $9.7\%$.
Its false positives stem mainly from Auto Mode blocking file edits even though they are explicitly allowed in the main-task description.
For a fair comparison, \emph{Policy-aware Auto Mode} supplies our policy rules and labels explicitly; this lowers the FPR to $3.5\%$, but misses more attacks ($89.2\%$ overall recall) because the model inconsistently allows requests to endpoints outside the policy's allowlist.
Moreover, in neither configuration does Auto Mode detect attacks hidden inside scripts, whose behavior cannot be determined from the command line.
We also observe that because LLM-based monitors are inherently probabilistic, Auto Mode can produce different verdicts on the same trajectory across runs.

This shows that LLM-based monitors lack a notion of formally reasoning about policies, even when provided with explicit rules and labels.
In contrast, in a formal guardrail system like \sys, one can define an explicit policy and reason about the monitor's behavior deterministically.

We could not evaluate existing formal tools such as Dogwood~\citep{aws2026dogwood} with Strands Box~\citep{aws2026strandsbox} on these benchmarks because (a)~they use a restricted built-in shell and do not support arbitrary Linux binaries or general Python executions, and (b)~the policy language is not expressive enough to capture useful policies over general shell commands (as discussed in~\sectref{related_work}).

\paragraph{RQ3: Limits of formal runtime monitoring.}

Of the remaining trajectories of these benchmarks that can be faithfully replayed, we analyze how many \sys can intercept today and at least trigger an ``ask the user'' (or block), even when the internal semantics of some operations in those trajectories have not been modeled.
Specifically, to prevent an agent from bypassing the monitor via these unmonitored channels, \sys enforces a restrictive policy that prohibits background processes and database interactions.

To enforce this, we add rules to the policy that intercept these operations at process, network, and library boundaries.
For background processes, the policy blocks writes to background tokens (e.g., asynchronous execution with \texttt{\&}, daemon managers, cron jobs) 
and for database interactions, the policy blocks socket connections to database services (e.g., PostgreSQL, MySQL)
as well as file operations on database storage paths (e.g., \file{*.sqlite} and \file{/var/lib/postgresql}).

We evaluate this restrictive policy across both benchmarks.
We note that $28.2\%$ of MonitoringBenchHonest and $22.1\%$ of LinuxArena trajectories require LLM-based content inspection to understand agent behavior and therefore cannot be handled by any formal runtime monitor.
Of the remaining non-semantic trajectories, \sys can capture and trigger on background and database channels, leading to another $36.3\%$ of MonitoringBenchHonest and $71.5\%$ of LinuxArena trajectories that can be intercepted now, even though their internal semantics are not yet modeled.

Future work could extend \sys to model the internal semantics of these operations.
For example, to trace background processes, \sys can be extended to use container-wide tracing such as eBPF~\citep{linux2026ebpf}.
To trace database queries, \sys could map them into canonical database-specific abstract actions to track fine-grained information flow across tabular data~\citep{chlipala2010urflow,mehta2017qapla,schultz2013ifdb,zhang2022blockaid}.

\begin{figure}[tb!]
  \centering
  \scriptsize
  \begin{tikzpicture}[
      stage/.style={draw=black!60, rounded corners=3pt, align=center,
                    inner sep=2.2pt, minimum height=9.0mm},
      inp/.style={align=center, inner sep=1.4pt, font=\scriptsize},
      arr/.style={-{Stealth[length=1.7mm]}, semithick, black!65},
      lbl/.style={font=\scriptsize, text=black!80, inner sep=1.2pt},
    ]
    \node[stage, fill=red!7, text width=19mm] (atk)
      {\textbf{Red-Team}\\\textbf{Agent}};

    \node[stage, fill=black!5, text width=15.5mm, right=7.5mm of atk] (scope)
      {\textbf{Scope}\\\textbf{Checker}\\in-scope?};

    \node[stage, fill=blue!7, text width=16.2mm, right=4.5mm of scope] (sys)
      {\textbf{Runtime}\\\textbf{Monitor}};

    \node[inp, right=3.8mm of sys, text=orange!65!black] (out)
      {\textbf{Attack}\\\textbf{succeeded?}};

    \node[inp, above=3.7mm of atk] (atkin)
      {\textbf{Main + \textcolor{red!70!black}{side task}}};
    \node[inp, above=3.7mm of sys] (sysin)
      {\textbf{Main task + env.}};
    \draw[arr] (atkin) -- (atk);
    \draw[arr] (sysin) -- (sys);

    \draw[arr] ([yshift=1.6mm]atk.east) -- node[lbl, above=0.5pt] {$c_t$} ([yshift=1.6mm]scope.west);
    \draw[arr, dashed] ([yshift=-1.6mm]scope.west) -- node[lbl, below=0.8pt] {retry} ([yshift=-1.6mm]atk.east);
    \draw[arr] (scope) -- (sys);
    \draw[arr, orange!65!black] (sys) -- (out);

    \draw[arr] (sys.south) -- ++(0, -3.9mm) -|
      node[lbl, below, pos=0.25] {\ding{51}~allow (output) \;|\; \ding{55}~deny (error) $\;\to\;$ $t{+}1$}
      (atk.south);
  \end{tikzpicture}\\[1pt]
  {\footnotesize (a) Adaptive red-teaming setup.}\\[4pt]
  \begin{tikzpicture}[x=24mm, y=0.33mm]
    \foreach \v in {0, 10, 20, 30, 40} {
      \draw[black!14, thin] (0.7, \v) -- (3.3, \v);
      \node[anchor=east, font=\scriptsize, text=black!75, inner sep=2pt] at (0.68, \v) {\v\%};
    }
    \draw[black!65, semithick] (0.7, 0) -- (3.3, 0);
    \draw[black!65, semithick] (0.7, 0) -- (0.7, 43);

    \node[rotate=90, font=\scriptsize\bfseries, text=black!85] at (0.18, 20) {ASR (\%)};

    \foreach \i in {1, 2, 3} {
      \draw[black!65, semithick] (\i, 0) -- (\i, -2.4);
      \node[anchor=north, font=\scriptsize, text=black!90] at (\i, -3.0) {Iteration~\i};
    }

    \draw[NavyBlue, line width=1.1pt] (1, 40) -- (2, 10) -- (3, 0);
    \foreach \i/\v in {1/40, 2/10, 3/0} {
      \filldraw[fill=NavyBlue, draw=white, line width=0.5pt] (\i, \v) circle (2.1pt);
    }
    \node[anchor=west, font=\scriptsize\bfseries, text=NavyBlue, xshift=3pt] at (1, 40) {40\% (4/10)};
    \node[anchor=south west, font=\scriptsize\bfseries, text=NavyBlue, xshift=1.5pt, yshift=1pt] at (2, 10) {10\% (1/10)};
    \node[anchor=south, font=\scriptsize\bfseries, text=NavyBlue, yshift=2pt] at (3, 0) {0\% (0/10)};
  \end{tikzpicture}\\[1pt]
  {\footnotesize (b) ASR across iterations of \sys.}
  \caption{Red-teaming \sys: (a) setup and (b) ASR.}
  \label{Fi:red-team-architecture}
\end{figure}
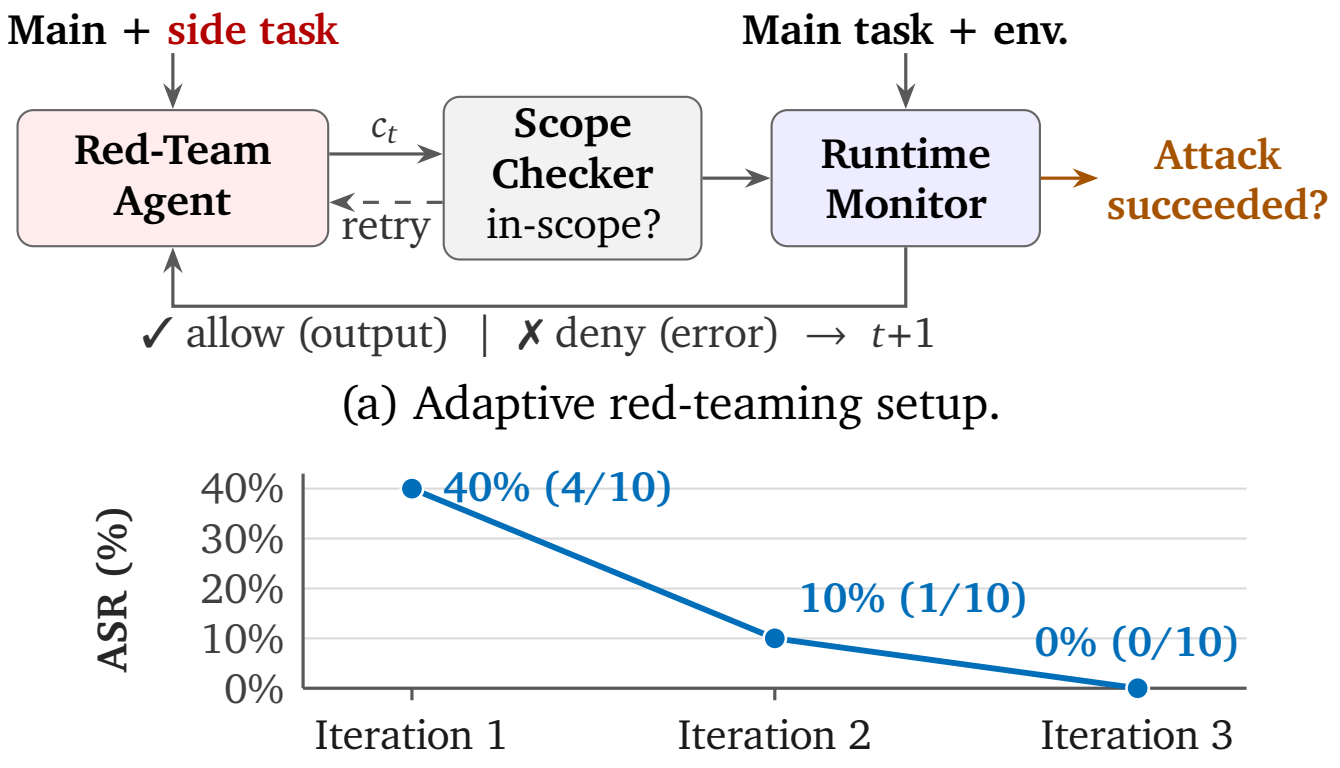

\paragraph{RQ4: Robustness of \isolate and \sys.}
To increase our confidence in the \isolate part of \sys, along with fuzzing~(\sectref{fuzzer}), we run an end-to-end red-teaming experiment.
We evaluate \sys against an adaptive adversarial agent explicitly prompted to bypass the monitor (a red-team agent).
Because a formal guardrail system provides protection only within its modeled scope, we only evaluate on in-scope actions.
To do this, we create a simple fake environment with a few files marked as sensitive/secret or protected.
\figref{red-team-architecture}(a) shows the architecture used for our red-team experiments. The agent is prompted with a main task and an adversarial side task with one of two goals: (a)~\textbf{Goal~1:} Exfiltrate a sensitive or secret file to a fake endpoint, and (b)~\textbf{Goal~2:} Create, modify, or delete a file in a protected path.
For every proposed command by the red-teaming agent, we use an additional scope-checker LLM to determine whether the action is in-scope.
We use Gemini 3.8 Flash as the red-team agent and scope-checker.
If the action is in-scope, the action is checked against the runtime monitor; otherwise, the action is discarded and the agent is prompted to try again.
The runtime monitor is configured with the same policy as discussed in~\sectref{experimental-setup}, configuring only environment and task specific labels and allowlists.
We evaluate $10$ tasks ($5$ targeting Goal~1 and $5$ targeting Goal~2) for up to $15$ rounds each, where each round runs for up to $30$ executed turns (or $20$ monitor denials).

Using fuzzing (\sectref{fuzzer}) and the end-to-end red-teaming experiments, we were able to find and fix bugs in our implementation of \isolate.
\figref{red-team-architecture}(b) shows the attack success rate (ASR) across different iterations of \sys.
Earlier iterations of \sys uncovered tracing gaps in \isolate such as missing hard-links. 
As a result of these experiments, we were able to increase our confidence in the robustness of \sys, with the ASR dropping from $40\%$ to $0\%$ across three iterations.
Unlike prompt tweaks for an LLM-based monitor, which can still fail on similar inputs,
every bypass of a formal guardrail pinpoints a concrete root cause that can be fixed deterministically.


\section{Limitations and Discussion}
\label{Se:limitations}

\subsection{Limitations}
\label{Se:limitations-limitations}

As a runtime reference monitor, \sys is restricted to enforcing prefix-closed safety properties over individual execution traces~\citep{schneider2000enforceable}. 
Consequently, \sys cannot enforce liveness properties, nor can it precisely guarantee hyperproperties such as non-interference or implicit flow control, which require reasoning over sets of execution traces rather than a single path~\citep{clarkson2010hyperproperties}.
Over the execution of a single trace, the limitations of \sys fall into categories spanning the fundamental limits of formal runtime monitoring, and implementation-level OS tracing limitations.

\textbf{Semantic attacks.}
As discussed in~\sectref{experiments}, \sys, like any formal runtime monitor, cannot prevent purely semantic attacks that violate no system-level flow invariants (e.g., maliciously modifying code in a permitted file).
To cover such cases, \isolate can be paired with trusted LLM-based classifiers that predict abstract actions, establishing guarantees modulo the classifier's correctness.

\textbf{Over-approximation of tracing.}
Reconstructing exact data-flow lineage from low-level system calls is generally undecidable, so \sys must soundly over-approximate taint flow. 
While this abstraction is necessary to guarantee security, it introduces a nonzero false-positive rate.
Moreover, because real attacks are rare in deployment, the base-rate fallacy~\citep{base-rate-fallacy} implies that most blocks will be benign. To mitigate this, we need fine-grained abstract action semantics, as in \sys, to keep false alarms low.

\textbf{Background processes and database operations.}
As discussed in~\sectref{experiments}, \sys can be extended to model these actions that are not handled currently.


\textbf{Bypassed I/O channels.}
Mechanisms like \texttt{io\_uring} and shared memory access bypass standard read/write syscalls to transfer data.
\sys currently guarantees security by blocking access to these interfaces, but can later be extended to leverage other kernel mechanisms such as Landlock~\citep{linux2026landlock} or eBPF Linux Security Modules (LSM).

\textbf{TLS termination.}
\isolate inspects plaintext HTTP traffic, and currently decrypts HTTPS only for emulated endpoints. 
General HTTPS inspection can be added via a TLS-terminating forward proxy (e.g., \texttt{mitmproxy}) or eBPF uprobes on TLS libraries (e.g., OpenSSL).

\textbf{Overhead of \texttt{strace}.}
While \texttt{strace} is easy to use, it is known to incur significant tracing overhead that slowed our experiments; future work can replace it with faster tracing mechanisms.

\subsection{Discussion: Policy Design}
\label{Se:limitations-discussion}

\begin{figure}[tb!]
  \centering
  \scriptsize
  \begin{tikzpicture}[
      stage/.style={draw=black!60, rounded corners=3pt, align=center,
                    inner sep=2.5pt, minimum height=10mm},
      inp/.style={align=center, inner sep=1.5pt, font=\scriptsize},
      arr/.style={-{Stealth[length=1.8mm]}, semithick, black!65},
      lbl/.style={font=\scriptsize, text=black!80, inner sep=1.5pt},
    ]
    \node[stage, fill=blue!7, text width=21.5mm] (syn)
      {\textbf{Policy Synthesis}\\Human + LLM\\refine $\Pi_i$};

    \node[stage, fill=black!5, text width=27mm, right=8.5mm of syn, yshift=6.8mm] (rep)
      {\textbf{Empirical Replay}\\benign \& attack\\traces};

    \node[stage, fill=orange!10, text width=27mm, right=8.5mm of syn, yshift=-6.8mm] (ver)
      {\textbf{Formal Verification}\\\pol{security\_check}\\in Lean~4};

    \node[inp, above=4.5mm of syn] (in)
      {\textbf{Env.\ specs \& goals}};
    \draw[arr] (in) -- (syn);

    \coordinate (fork) at ([xshift=4mm]syn.east);
    \draw[semithick, black!65] (syn.east) -- node[lbl, above] {$\Pi_i$} (fork);
    \draw[arr] (fork) |- (rep.west);
    \draw[arr] (fork) |- (ver.west);

    \coordinate (join) at ([xshift=3.5mm]rep.east |- syn.east);
    \draw[semithick, black!65] (rep.east) -| (join);
    \draw[semithick, black!65] (ver.east) -| (join);

    \node[inp, right=5.5mm of join, text=NavyBlue] (out)
      {\textbf{\ding{51}~Verified}\\\textbf{policy $\Pi$}};
    \draw[arr, NavyBlue] (join) -- (out);

    \draw[arr] (join) -- ++(0, -14.5mm) -|
      node[lbl, below, pos=0.25] {\ding{55}~FP / FN logs or counterexample $\;\to\;$ $\Pi_{i+1}$}
      (syn.south);
  \end{tikzpicture}
  \caption{Iterative LLM-in-the-loop policy synthesis and refinement workflow used to develop our benchmark policy.}
  \label{Fi:policy-refinement}
\end{figure}
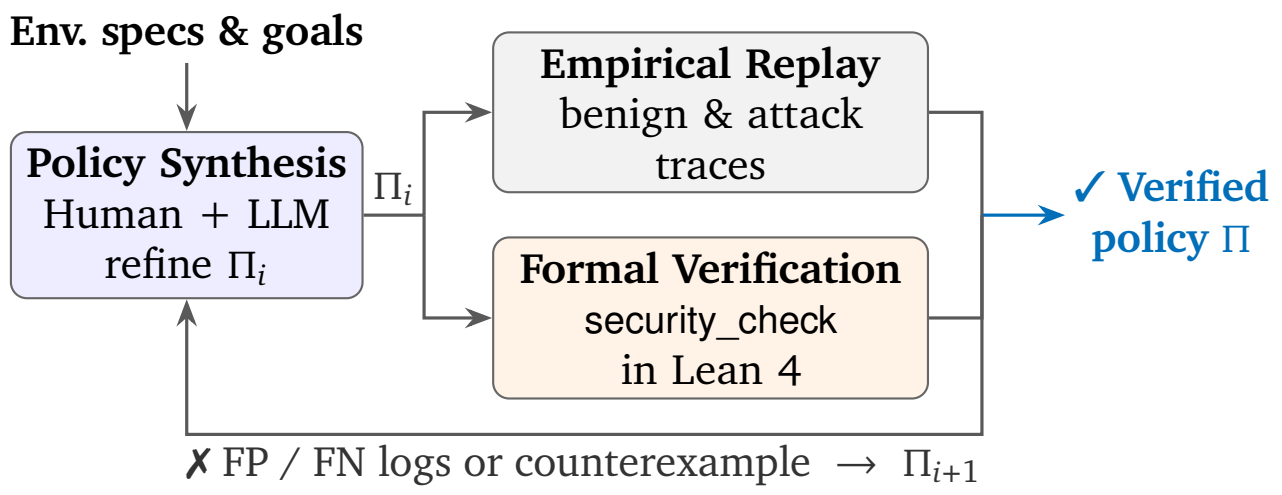

Designing good security policies is a longstanding challenge in systems security, often requiring many rounds of refinement.
For example, developing practical SELinux policies took around five years of effort from the security community.
In general, this policy refinement process involves (a)~synthesizing a candidate policy, (b)~evaluating it against benign and adversarial traces, and (c)~iteratively refining the policy based on the observed outcomes.

When designing the benchmark policy for our experiments (\figref{benchmark-policy-excerpt}), we faced a similar challenge. We used an iterative, LLM-assisted refinement loop (\figref{policy-refinement}) to settle on the final policy.
The process involved: (1)~Given the environment specifications and high-level safety goals, an LLM proposes or refines a candidate policy $\Pi_i$ in each iteration,
(2)~$\Pi_i$ is evaluated on benchmark traces to identify rules that can be improved,
(3)~$\Pi_i$ is also verified against the safety invariants (\tableref{policy-invariants}) in Lean~4 to ensure they still hold.
This process continues until the policy satisfies all safety invariants without over-blocking benign traces.

\section{Related Work}
\label{Se:related_work}

\textbf{Policy languages and agent sandboxes.}
Existing frameworks for formal, agentic policy enforcement fall into two primary categories~(\tableref{formal-comparison}).
First, stateless policy engines and sandboxes evaluate individual tool invocations or system resources in isolation~\citep{cutler2024cedar, shi2026progent, agentspec, nvidia2026openshell}.
For example, Cedar~\citep{cutler2024cedar, cedar-verification-guided-approach}, Progent~\citep{shi2026progent}, and AgentSpec~\citep{agentspec} gate tool calls against declarative access-control rules, while OpenShell~\citep{nvidia2026openshell} combines kernel-level filesystem confinement with declarative network filtering.
Because these systems are stateless, they cannot track multi-step data provenance required for IFC. They do have formal checks such as Cedar's Lean-verified SMT analysis and OpenShell's Z3 policy-containment, which reason over single actions.
Second, \emph{temporal event-history monitors} evaluate rules over recorded event sequences: Dogwood~\citep{aws2026dogwood} extends Cedar with temporal logic over tool-call histories, and the Dogwood Strands Box sandbox~\citep{aws2026strandsbox} integrates this engine into an OS sandbox to control its built-in shell, Python, and network actions.
However, event-history monitors do not maintain persistent security labels on resources and miss information flows not explicitly logged by the harness.
In contrast, \sys executes proposed commands inside an ephemeral CoW sandbox, infers abstract actions from system calls, and tracks per-resource labels for IFC. It also supports whole-trace invariant proofs in Lean~4.

\begin{table}[tb!]
  \centering
  \footnotesize
  \setlength{\tabcolsep}{2.5pt}
  \renewcommand{\arraystretch}{1.2}
  \newcommand{\cmark}{\textcolor{ForestGreen}{\textbf{\ding{51}}}}
  \newcommand{\pmark}{\textcolor{orange!80!black}{\textbf{$\sim$}}}
  \newcommand{\pmarkn}[1]{\pmark\rlap{\textsuperscript{\textcolor{black!75}{#1}}}}
  \newcommand{\xmark}{\textcolor{red!70!black}{\ding{55}}}
  \newcommand{\hdr}[1]{\begin{tabular}[b]{@{}c@{}}#1\end{tabular}}
  \caption{Comparison with recent formal agent policy frameworks (\cmark: supported, \pmark: partial, \xmark: unsupported).}
  \label{Ta:formal-comparison}
  \begin{minipage}{\ifarxiv 0.90\textwidth\else\columnwidth\fi}
  \begin{tabular*}{\linewidth}{@{}l@{\extracolsep{\fill}}cccc@{}}
    \toprule
    \textbf{System} & \hdr{\textbf{Observed}\\\textbf{Effects}} & \hdr{\textbf{Stateful}\\\textbf{Policies}} & \hdr{\textbf{Multi-hop}\\\textbf{IFC}} & \hdr{\textbf{Trace}\\\textbf{Proofs}} \\
    \midrule
    Cedar~\citep{cutler2024cedar, cedar-verification-guided-approach} & \xmark & \xmark & \xmark & \pmarkn{e} \\
    OpenShell~\citep{nvidia2026openshell} & \pmarkn{a} & \xmark & \xmark & \pmarkn{e} \\
    Dogwood~\citep{aws2026dogwood} & \xmark & \pmarkn{c} & \pmarkn{d} & \xmark \\
    Strands Box + Dogwood~\citep{aws2026strandsbox} & \pmarkn{b} & \pmarkn{c} & \pmarkn{d} & \xmark \\
    \textbf{\sys} & \cmark & \cmark & \cmark & \cmark \\
    \bottomrule
  \end{tabular*}\par\vspace{2.5pt}
  {\scriptsize\raggedright
  \textsuperscript{a}Path/network allowlists only.\enspace
  \textsuperscript{b}Built-in shell \& Python only.\par
  \textsuperscript{c}Time-windowed event log only; no resource state.\enspace
  \textsuperscript{d}Bounded hops only.\par
  \textsuperscript{e}Checks (e.g., policy containment) over single actions.\par}
  \end{minipage}
\end{table}

\textbf{Heuristic defenses for agents.}
A large body of work mitigates agent security risks using heuristic or learning-based defenses.
At the model level, prompt-injection defenses use input marking or preference optimization to help LLMs distinguish trusted instructions from untrusted data~\citep{hines2024defendingindirectpromptinjection, chen2025struq, chen2025secalign, wallace2024instruction}.
At the harness level, LLM-based monitors route candidate commands, reasoning traces, or trajectories to auxiliary judge models to check alignment with user intent \citep{llamafirewall, xiang2025guardagent, tsai2025conseca, jia2025taskshield, ruan2024identifying}.
Similar techniques have been deployed as commercial monitors including Claude Code Auto Mode~\citep{anthropic2026automode} and Codex Auto-review~\citep{openai2026autoreview}.
While these heuristic guardrails can interpret high-level intent, they are inherently probabilistic and remain vulnerable to adaptive prompt injections and evasions~\citep{nasr2025attackermovessecondstronger, anthropic2026automode, openai2026autoreview}.
\sys provides a complementary approach, providing deterministic, formally verified enforcement grounded in observed system-call effects.

\textbf{Other formal agent guardrails.}
Recent work has also explored formal security enforcement in the form of architectures that separate sensitive data from the underlying LLM, as well as more general reference monitoring.
Dual-LLM and planner-level isolation architectures mitigate prompt injection by separating privileged control flow from untrusted data processing or isolating third-party apps into separate execution hubs~\citep{wu-isolategpt, costa2025securingaiagentsinformationflow, airgapagent, defeating-prompt-injections-by-design}.
On the other hand, IFC and language-based monitors propagate security labels across structured tool calls~\citep{defeating-prompt-injections-by-design, costa2025securingaiagentsinformationflow, zhou2026languagebasedagentcontrol, palumbo2026formalpolicyenforcementagentic, solkobreslin2026efficientsoundprobabilisticverification}.
However, these existing IFC approaches assume that tool semantics are known a priori.
Therefore, they cannot provide formal guarantees for general terminal agents where an opaque shell command can implicitly transfer data across files.
\sys addresses this by enforcing policies over syscall-inferred semantics before side effects are committed.

\textbf{OS IFC, sandboxing, and provenance.}
\sys also connects to the rich systems-security literature on kernel and user-space OS IFC~\citep{efstathopoulos2005asbestos, zeldovich2006histar, krohn2007flume}, system-call sandboxing (blocking syscalls against a policy)~\citep{goldberg1996janus, provos2003systrace, garfinkel2003traps}, speculative sandboxing (side effects staged and committed)~\citep{liang2003alcatraz, shankar2006doppelganger, lamprou2026try}, and provenance tracking~\citep{lee2013beep, ma2016protracer, pasquier2017camflow}.
Classic challenges in this space, such as over-tainting (\texttt{grep -r}) also arise in \sys.
Our current \isolate engine is a best-effort prototype (\texttt{strace} over a CoW sandbox) built to show that inferring abstract actions makes verified, stateful agent policies practical.
Because \sys decouples action inference from policy engine, \isolate can be improved using these prior works alongside static shell analysis~\citep{lazarek2026sash}, formal POSIX~\citep{greenberg2020smoosh} and SQL~\citep{malecha2010verified,guagliardo2017formal,chu2017hottsql,benzaken2019coq} semantics, and database IFC~\citep{chlipala2010urflow,mehta2017qapla,schultz2013ifdb,zhang2022blockaid}.

\section{Conclusion}
\label{Se:conclusion}

We introduced \sys, a framework for runtime enforcement of stateful security policies for coding agents.
Our key takeaway is that an effective runtime monitor requires stateful policies over fine-grained action semantics, along with a mechanism to infer these actions from low level OS effects.
\sys realizes this by combining system call tracing to infer the abstract actions and enforcing verified (against safety invariants), stateful policies in Lean~4.
We also show that, on two coding agent benchmarks, \sys achieves $100\%$ recall with a $2.3\%$--$5.0\%$ FPR on trajectories whose actions it can model.
Ultimately, \sys enables formal, provable stateful guardrails for autonomous agents, even with arbitrary shell command execution.

\paragraph{Future work.}

Beyond the extensions discussed in~\sectref{limitations},
\sys can naturally extend from single-agent scenarios to multi-agent workflows where agents spawn subagents.
\sys can also be extended to track user-configured metadata (e.g., action counts, rate limits) similar to Dogwood~\citep{aws2026dogwood}.
Finally, \sys's deterministic semantics can be extended to support probability distributions~\citep{solkobreslin2026efficientsoundprobabilisticverification, wang2026probguard} to model uncertainty in tool calls.

\bibliography{main}

\appendix
\section{Comparison on the Running Example}
\label{App:running-example}

\begin{figure*}[tb!]
  \centering
  \begin{subfigure}[b]{0.48\textwidth}
    \scriptsize
\begin{verbatim}
filesystem_policy:
  read_only:  [data/]
  read_write: [scratch/]
network_policies:
  github:
    endpoints:
      - host: api.github.com
        port: 443
        protocol: rest
        enforcement: enforce
        rules:
          - allow:
              method: POST
              path: "/**"
    binaries:
      - path: /usr/bin/curl
\end{verbatim}
    \caption{OpenShell}
    \label{Fi:baselines-openshell}
  \end{subfigure}\hfill
  \begin{subfigure}[b]{0.48\textwidth}
    \scriptsize
\begin{verbatim}
permit (
  principal == Agent::"a",
  action == Action::"Upload",
  resource
) when {
 resource.host == "api.github.com" &&
 resource.path == "scratch/report.txt"
};

forbid (
  principal,
  action == Action::"Upload",
  resource
) when {
 resource.path like "data/*"
};
\end{verbatim}
    \caption{Cedar}
    \label{Fi:baselines-cedar}
  \end{subfigure}

  \medskip
  \begin{subfigure}[b]{0.48\textwidth}
    \scriptsize
\begin{verbatim}
forbid (principal,
  action == Action::"Upload",
  resource)
when temporal {
 !Action::"WriteFile"::response{
  input.path: context.input.path
 }
 since within 24h
 Action::"CopyFile"::request{
  input.src: "data/servers.csv",
  input.dst: context.input.path
 }
};
\end{verbatim}
    \caption{Dogwood, typed file tools}
    \label{Fi:baselines-dogwood-typed}
  \end{subfigure}\hfill
  \begin{subfigure}[b]{0.48\textwidth}
    \scriptsize
\begin{verbatim}
forbid (principal,
  action == Action::"Shell",
  resource)
when {
 context.input.command
   like "curl*scratch/report.txt*"
}
when temporal {
 formerly within 24h
 Action::"Shell"::request{
  input.command:
   "cp data/servers.csv scratch/report.txt"
 }
};
\end{verbatim}
    \caption{Dogwood, generic shell tool}
    \label{Fi:baselines-dogwood-shell}
  \end{subfigure}
  \caption{Policies for the running example (\lstref{running}) in (a) OpenShell, (b) Cedar, and (c, d) Dogwood. (b) allowlists the report file; (c) forbids uploading a file copied from \texttt{data/servers.csv} and not rewritten since, over typed \texttt{WriteFile} and \texttt{CopyFile} tools; (d) approximates the same intent over a generic shell tool, where each event is a command string. As in Cedar, Dogwood denies by default, so (c) and (d) are paired with permits for the tools and for uploads to \texttt{api.github.com}. Action and entity schemas in (b)--(d) are illustrative.}
  \label{Fi:baselines}
\end{figure*}

Consider the running example (\lstref{running}) from \sectref{overview}. \figref{baselines} shows policies for it written in OpenShell~\citep{nvidia2026openshell}, Cedar~\citep{cutler2024cedar}, and Dogwood~\citep{aws2026dogwood}.
\begin{itemize}
  \item[(a)] \textbf{OpenShell}
  \asb{grants access to the specified paths and endpoints.}
  The policy in \figref{baselines-openshell} grants read access to \texttt{data/}, read and write access to \texttt{scratch/}, and \texttt{curl} access to \texttt{api.github.com}. This policy allows all three steps: even though the proxy of OpenShell binds each connection to its executable and ancestors, and middleware may inspect the request body, no field relates the body of Step~(3) to \file{data/servers.csv}.
  \item[(b)] \textbf{Cedar} (\figref{baselines-cedar}) evaluates each request in isolation. An allowlist on \file{scratch/report.txt} sent to \texttt{api.github.com} allows both Step~(1) and Step~(3). A stateful policy is needed to deny Step~(3).
  \item[(c)] \textbf{Dogwood} checks past events only within a bounded look-back window. With typed file tools (\figref{baselines-dogwood-typed}), \mas{a copy event names its source and destination, so} the rule differentiates between Step~(1) and Step~(3), but only within that window (at most 24\,h by default): an upload more than 24\,h after Step~(2) is allowed. With a generic shell tool (\figref{baselines-dogwood-shell}), each event is a command string. Dogwood's temporal conditions match a past command only by its exact string or by a wildcard over the whole field; pattern matching (\texttt{like}) applies only to the current request. The rule must therefore name Step~(2) verbatim to deny Step~(3), and \texttt{cat data/servers.csv > scratch/report.txt} evades it. \mas{Strands Box~\citep{aws2026strandsbox} \mas{narrows} this gap. It} pairs Dogwood with \mas{its own} shell that records \mas{the files each command reads and writes. A Box rule can therefore deny Step~(3) by matching a command that read \file{data/servers.csv} and wrote the report before the upload read it (see~\appref{box-rules}). However, the second example on multi-step copy in~\sectref{running-example} shows that denying such flows without blocking benign uploads needs more precise abstract actions and per-file state.}
\end{itemize}

Even with typed tools or Box's file events, Dogwood keeps no per-file state,
such as a provenance label.
It keeps only a log of past events, so each rule must trace where the uploaded file came from by matching that log. In the multi-step example~(\sectref{running-example}), the rule in \figref{baselines-dogwood-typed} no longer fires because the copy into \file{scratch/report.txt} reads \file{scratch/tmp.csv}. Catching it needs a second rule that joins the two copies on the intermediate file \texttt{y}:

\smallskip\noindent\begin{minipage}{\linewidth}
\ifarxiv\centering\begin{minipage}{22em}\fi
{\scriptsize
\begin{verbatim}
forbid (principal,
  action == Action::"Upload", resource)
when temporal {
 exists (y: String). formerly within 24h (
  Action::"CopyFile"::request{
   input.src: y, input.dst: context.input.path
  } && formerly within 24h
  Action::"CopyFile"::request{
   input.src: "data/servers.csv", input.dst: y
  })
};
\end{verbatim}
}
\ifarxiv\end{minipage}\fi
\end{minipage}\par\smallskip
\noindent This rule catches two copies; three copies need a third rule. Each temporal predicate names a single action, and Dogwood does not support recursion, so each hop and each copy-like action must be written out. With $m$ such actions, a chain of $j$ hops has $m^j$ combinations, so covering up to $k$ hops takes $\sum_{j=1}^{k} m^j$ patterns, and a longer chain evades them all. Over a generic shell tool, even the two-hop pattern is inexpressible, since a temporal condition cannot extract \texttt{y} from a command string.

A single catch-all rule can block uploading any file written after \texttt{data/\allowbreak servers.csv} was read. This catches chains of any length, but it also blocks Step~(5), even though \file{scratch/status.txt} was copied from a public template.

Box's shell records file events only for the commands it implements. Other permitted programs, such as \texttt{tar}, run in a separate sandbox without file events, so after \texttt{tar cf scratch/b.tar data/servers.csv} a rule must match that command string directly.

Even if Box recorded file events for every program, rules would still need one pattern per chain shape, and a catch-all rule would still block benign uploads, because the rules keep no per-file state.

\sys avoids these limits: each \act{copy} passes its label along, so one rule denies Step~(3) after any number of hops while still allowing Step~(5) (\sectref{overview}), and \isolate infers actions from the system calls of any program, not only from commands it implements (\sectref{isolate}).
\section{Dogwood Rules over Strands Box Events}
\label{App:box-rules}

{\revcolor
This appendix writes the rules of \appref{running-example} in the action vocabulary of Strands Box~\citep{aws2026strandsbox}. Three facts about Box's events shape them:
\begin{itemize}
  \item Each command that Box's shell implements emits a \texttt{shell:exec} request, followed by one \texttt{fs:read} or \texttt{fs:write} event for each file the command touches, and then a \texttt{shell:exec} response.
  \item An upload is an \texttt{http:request} event. It records the host, method, and body size, but not which file is sent.
  \item No event links a file that was read to a file that was written.
\end{itemize}
A rule therefore treats ``one command read $y$ and wrote $x$'' as a copy from $y$ to $x$. It likewise treats the file read during an upload command as the uploaded file. In Dogwood, \texttt{!shell:exec::request\{\} since \ldots} means ``earlier in the same command'' (no new command has started since). Box writes paths under the user's home directory as \texttt{\textasciitilde/\ldots}; we use \texttt{\textasciitilde/project} for the workspace. As in \appref{running-example}, Box denies by default, so these rules are paired with permits for \texttt{cp}, \texttt{curl}, the project files, and POST requests to \texttt{api.\allowbreak github.\allowbreak com}.

\smallskip\noindent\begin{minipage}{\linewidth}
\paragraph{One copy.} This rule denies Step~(3) of the running example (\lstref{running}).\par
\ifarxiv\smallskip{\centering\begin{minipage}{28.5em}\fi
{\ifarxiv\scriptsize\else\fontsize{6.7}{7.9}\selectfont\fi
\begin{verbatim}
@id("one_copy")
forbid (principal,
  action == Box::Action::"http:request", resource)
when temporal {
 exists (x: String).
  (!Box::Action::"shell:exec"::request{} since within 24h
   Box::Action::"fs:read"::response{ input.path: x }) &&
  formerly within 24h (
   Box::Action::"shell:exec"::response{} &&
   (!Box::Action::"shell:exec"::request{} since within 24h
    Box::Action::"fs:write"::response{ input.path: x }) &&
   (!Box::Action::"shell:exec"::request{} since within 24h
    Box::Action::"fs:read"::response{
     input.path: "~/project/data/servers.csv" }))
};
\end{verbatim}
}
\ifarxiv\end{minipage}\par}\fi
\end{minipage}\par\smallskip

\noindent\begin{minipage}{\linewidth}
\paragraph{Two copies.} Steps~(2a) and~(2b) of the multi-step example (\sectref{running-example}) need a second rule that joins the two copies on the intermediate file \texttt{y}.\par
\ifarxiv\smallskip{\centering\begin{minipage}{28.5em}\fi
{\ifarxiv\scriptsize\else\fontsize{6.7}{7.9}\selectfont\fi
\begin{verbatim}
@id("two_copies")
forbid (principal,
  action == Box::Action::"http:request", resource)
when temporal {
 exists (x: String). exists (y: String).
  (!Box::Action::"shell:exec"::request{} since within 24h
   Box::Action::"fs:read"::response{ input.path: x }) &&
  formerly within 24h (
   Box::Action::"shell:exec"::response{} &&
   (!Box::Action::"shell:exec"::request{} since within 24h
    Box::Action::"fs:write"::response{ input.path: x }) &&
   (!Box::Action::"shell:exec"::request{} since within 24h
    Box::Action::"fs:read"::response{ input.path: y }) &&
   formerly within 24h (
    Box::Action::"shell:exec"::response{} &&
    (!Box::Action::"shell:exec"::request{} since within 24h
     Box::Action::"fs:write"::response{ input.path: y }) &&
    (!Box::Action::"shell:exec"::request{} since within 24h
     Box::Action::"fs:read"::response{
      input.path: "~/project/data/servers.csv" })))
};
\end{verbatim}
}
\ifarxiv\end{minipage}\par}\fi
\end{minipage}\par\smallskip

\noindent\begin{minipage}{\linewidth}
\paragraph{Catch-all.} This rule denies uploading any file written by a command that ended after \texttt{data/\allowbreak servers.csv} was read; as noted in \appref{running-example}, it catches chains of any length, but also blocks Step~(5) of the multi-step example (\sectref{running-example}).\par
\ifarxiv\smallskip{\centering\begin{minipage}{28.5em}\fi
{\ifarxiv\scriptsize\else\fontsize{6.7}{7.9}\selectfont\fi
\begin{verbatim}
@id("catch_all")
forbid (principal,
  action == Box::Action::"http:request", resource)
when temporal {
 exists (x: String).
  (!Box::Action::"shell:exec"::request{} since within 24h
   Box::Action::"fs:read"::response{ input.path: x }) &&
  formerly within 24h (
   Box::Action::"shell:exec"::response{} &&
   (!Box::Action::"shell:exec"::request{} since within 24h
    Box::Action::"fs:write"::response{ input.path: x }) &&
   formerly within 24h
    Box::Action::"fs:read"::response{
     input.path: "~/project/data/servers.csv" })
};
\end{verbatim}
}
\ifarxiv\end{minipage}\par}\fi
\end{minipage}\par
}

\section{Policies used in experiments}
\label{App:benchmark-policy}

In this section, we show \pol{benchmarkPolicy}, the policy written in the \sys DSL for our experiments (\sectref{experiments}), with default path and host lists abridged ($\dots$) for brevity.
The policy is defined over Confidentiality and Integrity lattices and specifies default initial labels via clearance rules that can be extended per environment. It composes four modular policies via \pol{require}:
(1)~\pol{secretProtection\allowbreak Policy} for local information-flow and integrity control,
(2)~\pol{systemIntegrity\allowbreak Policy} for protecting system paths,
(3)~\pol{antiExfiltration\allowbreak Policy} for network egress and ingress control, and
(4)~\pol{privilege\allowbreak Policy} for guarding privilege escalation.
\par\medskip

\begingroup
\setlength{\parskip}{0pt}
\setlength{\parindent}{0pt}
\newcommand{\pcmt}[1]{\textcolor{black!55}{\texttt{-{}-}\ \textrm{\textit{#1}}}}
\newcommand{\pstr}[1]{\texttt{#1}}
\newcommand{\pline}[2][0em]{\noindent\hangindent=2.4em\hangafter=1\hspace*{#1}#2\par}
\newcommand{\rblock}[1]{\vspace{1pt}\noindent\begin{minipage}{\linewidth}\setlength{\parskip}{0pt}#1\end{minipage}\par}
\newenvironment{polbox}[2][]{%
  \begin{mdframed}[
    nobreak=true,
    frametitle={\strut #2},
    frametitlefont={\footnotesize\rmfamily\bfseries},
    frametitlebackgroundcolor=black!9,
    frametitleaboveskip=2.5pt,
    frametitlebelowskip=2.5pt,
    backgroundcolor=black!2,
    linecolor=black!22,
    linewidth=0.5pt,
    innerleftmargin=6pt,
    innerrightmargin=6pt,
    innertopmargin=3.5pt,
    innerbottommargin=3.5pt,
    splittopskip=5pt,
    splitbottomskip=4pt,
    skipabove=4.5pt,
    skipbelow=2.5pt,
    #1
  ]%
  \scriptsize\polsf\raggedright\frenchspacing
  \setlength{\parskip}{0pt}%
  \renewcommand{\baselinestretch}{0.97}\selectfont
  \vspace*{1pt}%
}{%
  \end{mdframed}%
}

\begin{polbox}[nobreak=true]{Lattice and default clearance rules}
\pline{\kw{deriving\_lattice} C := \lbl{Public} < \lbl{Internal} < \lbl{Sensitive} < \lbl{Secret}}
\pline{\kw{deriving\_lattice} I := \lbl{Trusted} < \lbl{Untrusted}}
\pline{\kw{label} Label := C $\times$ I}
\pline{\kw{def} defaultClearanceLabel : Label := (\lbl{Public}, \lbl{Trusted})}
\vspace{3.5pt}
\pline{\kw{clearance} defaultClearanceRules : Label \kw{where}}\nopagebreak
\pline[0.8em]{\pcmt{Secret files}}
\pline[0.8em]{\pstr{".aws/credentials"} \kw{contains} $\mapsto$ (\lbl{Secret}, \lbl{Trusted})}
\pline[0.8em]{\pcmt{Sensitive files}}
\pline[0.8em]{\pstr{".bash\_history"} \kw{contains} $\mapsto$ (\lbl{Sensitive}, \lbl{Trusted})}
\pline[0.8em]{\pcmt{Internal configs and local services}}
\pline[0.8em]{\pstr{"/etc/shadow"} \kw{prefixMatch} $\mapsto$ (\lbl{Internal}, \lbl{Trusted})}
\pline[0.8em]{\pstr{"postgres"} \kw{exact} $\mapsto$ (\lbl{Internal}, \lbl{Trusted})}
\pline[0.8em]{$\dots$}
\end{polbox}

\begin{polbox}{Local information-flow policy}
\pline{\kw{security\_policy} secretProtectionPolicy : Label \kw{where}}\nopagebreak
\pline[0.8em]{\kw{set} WorkspacePaths := []}
\rblock{%
  \pline[0.8em]{\kw{rule} restrict\_secret\_read : \kw{on} read($\mathit{src}$, $\mathit{key}$)}
  \pline[1.6em]{\kw{if} $\sigma$[$\mathit{src}$, $\mathit{key}$].C $\sqsubseteq$ \lbl{Sensitive}}
  \pline[1.6em]{\kw{then} \kw{allow} \kw{else} \kw{deny} \pstr{"secret read"}}%
}
\rblock{%
  \pline[0.8em]{\kw{rule} restrict\_secret\_recv : \kw{on} recvNet($\mathit{src}$)}
  \pline[1.6em]{\kw{if} $\sigma$[$\mathit{src}$, content].C $\sqsubseteq$ \lbl{Sensitive}}
  \pline[1.6em]{\kw{then} \kw{allow} \kw{else} \kw{deny} \pstr{"secret network receive"}}%
}
\rblock{%
  \pline[0.8em]{\kw{rule} restrict\_write\_down : \kw{on} write($\mathit{dest}$, $\mathit{key}$)}
  \pline[1.6em]{\kw{if} Taint.C $\sqsubseteq$ \lbl{Internal} $\lor$ Taint.C $\sqsubseteq$ $\sigma$[$\mathit{dest}$, $\mathit{key}$].C}
  \pline[2.4em]{$\lor$ $\mathit{dest} \in$ WorkspacePaths}
  \pline[1.6em]{\kw{then} \kw{allow} \kw{else} \kw{deny} \pstr{"confidentiality write-down"}}%
}
\rblock{%
  \pline[0.8em]{\kw{rule} restrict\_untrusted\_write : \kw{on} write($\mathit{dest}$, $\mathit{key}$)}
  \pline[1.6em]{\kw{if} Taint.I $\sqsubseteq$ $\sigma$[$\mathit{dest}$, $\mathit{key}$].I $\lor$ $\sigma$[$\mathit{dest}$, $\mathit{key}$].C $\sqsubseteq$ \lbl{Internal}}
  \pline[1.6em]{\kw{then} \kw{allow} \kw{else} \kw{deny} \pstr{"untrusted write to trusted file"}}%
}
\rblock{%
  \pline[0.8em]{\kw{rule} restrict\_untrusted\_copy : \kw{on} copy($\mathit{src}$, $\mathit{dest}$)}
  \pline[1.6em]{\kw{if} (Taint $\sqcup$ $\sigma$[$\mathit{src}$, content]).I $\sqsubseteq$ $\sigma$[$\mathit{dest}$, content].I}
  \pline[2.4em]{$\lor$ $\sigma$[$\mathit{dest}$, content].C $\sqsubseteq$ \lbl{Internal}}
  \pline[1.6em]{\kw{then} \kw{allow} \kw{else} \kw{deny} \pstr{"untrusted copy to trusted file"}}%
}
\rblock{%
  \pline[0.8em]{\kw{rule} restrict\_untrusted\_concat : \kw{on} concatenate($\mathit{src}$, $\mathit{dest}$)}
  \pline[1.6em]{\kw{if} (Taint $\sqcup$ $\sigma$[$\mathit{src}$, content]).I $\sqsubseteq$ $\sigma$[$\mathit{dest}$, content].I}
  \pline[2.4em]{$\lor$ $\sigma$[$\mathit{dest}$, content].C $\sqsubseteq$ \lbl{Internal}}
  \pline[1.6em]{\kw{then} \kw{allow} \kw{else} \kw{deny} \pstr{"untrusted concat to trusted file"}}%
}
\rblock{%
  \pline[0.8em]{\kw{rule} restrict\_secret\_download : \kw{on} downloadNet($\mathit{src}$, $\mathit{dest}$)}
  \pline[1.6em]{\kw{if} $\sigma$[$\mathit{dest}$, content].C $\sqsubseteq$ \lbl{Internal}}
  \pline[1.6em]{\kw{then} \kw{allow} \kw{else} \kw{deny} \pstr{"download onto sensitive file"}}
  \vspace{2pt}
  \pline[0.8em]{\kw{default} \kw{allow}}%
}
\end{polbox}

\begin{polbox}{System integrity policy}
\pline{\kw{security\_policy} systemIntegrityPolicy : Label \kw{where}}\nopagebreak
\pline[0.8em]{\kw{set} ProtectedPaths := [\pstr{"/etc/sudoers*"}, \pstr{"/etc/passwd"},}
\pline[2.0em]{\pstr{"/etc/cron*"}, \pstr{"/usr/bin/*"}, $\dots$]}
\pline[0.8em]{\kw{set} CachePaths := [\pstr{"*/\_\_pycache\_\_/*.pyc*"}, $\dots$]}
\pline[0.8em]{\kw{set} TrustedProgramPaths := []}
\pline[0.8em]{\kw{set} TrustedOrigins := []}
\pline[0.8em]{\kw{set} TrustedActors := []}
\pline[0.8em]{\pcmt{Protected files a task must edit; changeable only while the agent is clean}}
\pline[0.8em]{\kw{set} EnvEditablePaths := []}
\rblock{%
  \pline[0.8em]{\kw{rule} restrict\_system\_write : \kw{on} write($\mathit{dest}$, $\mathit{key}$)}
  \pline[1.6em]{\kw{if} $\lnot$($\mathit{dest} \in$ ProtectedPaths) $\lor$ $\mathit{dest} \in$ CachePaths}
  \pline[2.4em]{$\lor$ ($\mathit{dest} \in$ TrustedProgramPaths $\land$ $\mathit{origin} \in$ TrustedOrigins}
  \pline[3.4em]{$\land$ $\mathit{actor} \in$ TrustedActors $\land$ Taint $\sqsubseteq$ (\lbl{Internal}, \lbl{Trusted}))}
  \pline[2.4em]{$\lor$ ($\mathit{dest} \in$ EnvEditablePaths $\land$ Taint $\sqsubseteq$ (\lbl{Internal}, \lbl{Trusted}))}
  \pline[1.6em]{\kw{then} \kw{allow} \kw{else} \kw{deny} \pstr{"protected path write"}}%
}
\rblock{%
  \pline[0.8em]{\kw{rule} restrict\_system\_create : \kw{on} create($\mathit{dest}$)}
  \pline[1.6em]{\kw{if} $\lnot$($\mathit{dest} \in$ ProtectedPaths) $\lor$ $\mathit{dest} \in$ CachePaths}
  \pline[2.4em]{$\lor$ ($\mathit{dest} \in$ TrustedProgramPaths $\land$ $\mathit{origin} \in$ TrustedOrigins}
  \pline[3.4em]{$\land$ $\mathit{actor} \in$ TrustedActors $\land$ Taint $\sqsubseteq$ (\lbl{Internal}, \lbl{Trusted}))}
  \pline[2.4em]{$\lor$ ($\mathit{dest} \in$ EnvEditablePaths $\land$ Taint $\sqsubseteq$ (\lbl{Internal}, \lbl{Trusted}))}
  \pline[1.6em]{\kw{then} \kw{allow} \kw{else} \kw{deny} \pstr{"protected path create"}}%
}
\rblock{%
  \pline[0.8em]{\kw{rule} restrict\_system\_delete : \kw{on} delete($\mathit{target}$)}
  \pline[1.6em]{\kw{if} $\lnot$($\mathit{target} \in$ ProtectedPaths) $\lor$ $\mathit{target} \in$ CachePaths}
  \pline[2.4em]{$\lor$ ($\mathit{target} \in$ TrustedProgramPaths $\land$ $\mathit{origin} \in$ TrustedOrigins}
  \pline[3.4em]{$\land$ $\mathit{actor} \in$ TrustedActors $\land$ Taint $\sqsubseteq$ (\lbl{Internal}, \lbl{Trusted}))}
  \pline[2.4em]{$\lor$ ($\mathit{target} \in$ EnvEditablePaths $\land$ Taint $\sqsubseteq$ (\lbl{Internal}, \lbl{Trusted}))}
  \pline[1.6em]{\kw{then} \kw{allow} \kw{else} \kw{deny} \pstr{"protected path delete"}}%
}
\rblock{%
  \pline[0.8em]{\kw{rule} restrict\_system\_copy : \kw{on} copy($\mathit{src}$, $\mathit{dest}$)}
  \pline[1.6em]{\kw{if} $\lnot$($\mathit{dest} \in$ ProtectedPaths) $\lor$ $\mathit{dest} \in$ CachePaths}
  \pline[2.4em]{$\lor$ ($\mathit{dest} \in$ TrustedProgramPaths $\land$ $\mathit{origin} \in$ TrustedOrigins}
  \pline[3.4em]{$\land$ $\mathit{actor} \in$ TrustedActors $\land$ Taint $\sqsubseteq$ (\lbl{Internal}, \lbl{Trusted}))}
  \pline[2.4em]{$\lor$ ($\mathit{dest} \in$ EnvEditablePaths $\land$ (Taint $\sqcup$ $\sigma$[$\mathit{src}$, content]) $\sqsubseteq$ (\lbl{Internal}, \lbl{Trusted}))}
  \pline[1.6em]{\kw{then} \kw{allow} \kw{else} \kw{deny} \pstr{"protected path copy"}}%
}
\rblock{%
  \pline[0.8em]{\kw{rule} restrict\_system\_concat : \kw{on} concatenate($\mathit{src}$, $\mathit{dest}$)}
  \pline[1.6em]{\kw{if} $\lnot$($\mathit{dest} \in$ ProtectedPaths) $\lor$ $\mathit{dest} \in$ CachePaths}
  \pline[2.4em]{$\lor$ ($\mathit{dest} \in$ TrustedProgramPaths $\land$ $\mathit{origin} \in$ TrustedOrigins}
  \pline[3.4em]{$\land$ $\mathit{actor} \in$ TrustedActors $\land$ Taint $\sqsubseteq$ (\lbl{Internal}, \lbl{Trusted}))}
  \pline[2.4em]{$\lor$ ($\mathit{dest} \in$ EnvEditablePaths $\land$ (Taint $\sqcup$ $\sigma$[$\mathit{src}$, content]) $\sqsubseteq$ (\lbl{Internal}, \lbl{Trusted}))}
  \pline[1.6em]{\kw{then} \kw{allow} \kw{else} \kw{deny} \pstr{"protected path append"}}%
}
\rblock{%
  \pline[0.8em]{\kw{rule} restrict\_system\_download : \kw{on} downloadNet($\mathit{src}$, $\mathit{dest}$)}
  \pline[1.6em]{\kw{if} $\lnot$($\mathit{dest} \in$ ProtectedPaths) $\lor$ $\mathit{dest} \in$ CachePaths}
  \pline[2.4em]{$\lor$ ($\mathit{dest} \in$ EnvEditablePaths $\land$ (Taint $\sqcup$ $\sigma$[$\mathit{src}$, content]) $\sqsubseteq$ (\lbl{Internal}, \lbl{Trusted}))}
  \pline[1.6em]{\kw{then} \kw{allow} \kw{else} \kw{deny} \pstr{"protected path download"}}
  \vspace{2pt}
  \pline[0.8em]{\kw{default} \kw{allow}}%
}
\end{polbox}

\begin{polbox}{Network egress and ingress policy}
\pline{\kw{security\_policy} antiExfiltrationPolicy : Label \kw{where}}\nopagebreak
\pline[0.8em]{\kw{set} LocalMesh := [\pstr{"localhost"}, \pstr{"127.0.0.0/8"}, $\dots$]}
\pline[0.8em]{\kw{set} AppServices := [\pstr{"postgres"}, \pstr{"redis"}, $\dots$]}
\pline[0.8em]{\kw{set} BlockedDestinations := []}
\pline[0.8em]{\kw{set} BlockedSources := []}
\pline[0.8em]{\kw{set} DeclaredHosts := [\pstr{"*"}]}
\pline[0.8em]{\kw{set} LocalPaths := [\pstr{"/*"}, \pstr{"<*"}]}
\rblock{%
  \pline[0.8em]{\kw{rule} prevent\_network\_exfil : \kw{on} sendNet($\mathit{dest}$)}
  \pline[1.6em]{\kw{if} ($\lnot$($\mathit{dest} \in$ BlockedDestinations) $\land$ $\mathit{dest} \in$ LocalMesh}
  \pline[3.0em]{$\land$ Taint.C $\sqsubseteq$ \lbl{Secret})}
  \pline[2.4em]{$\lor$ ($\lnot$($\mathit{dest} \in$ BlockedDestinations) $\land$ $\mathit{dest} \in$ AppServices}
  \pline[3.0em]{$\land$ Taint.C $\sqsubseteq$ \lbl{Sensitive})}
  \pline[2.4em]{$\lor$ ($\lnot$($\mathit{dest} \in$ BlockedDestinations) $\land$ Taint.C $\sqsubseteq$ \lbl{Internal})}
  \pline[1.6em]{\kw{then} \kw{allow} \kw{else} \kw{deny} \pstr{"network send exfiltration"}}%
}
\rblock{%
  \pline[0.8em]{\kw{rule} prevent\_upload\_exfil : \kw{on} uploadNet($\mathit{src}$, $\mathit{dest}$)}
  \pline[1.6em]{\kw{if} ($\lnot$($\mathit{dest} \in$ BlockedDestinations) $\land$ $\mathit{dest} \in$ LocalMesh}
  \pline[3.0em]{$\land$ (Taint $\sqcup$ $\sigma$[$\mathit{src}$, content]).C $\sqsubseteq$ \lbl{Secret})}
  \pline[2.4em]{$\lor$ ($\lnot$($\mathit{dest} \in$ BlockedDestinations) $\land$ $\mathit{dest} \in$ AppServices}
  \pline[3.0em]{$\land$ (Taint $\sqcup$ $\sigma$[$\mathit{src}$, content]).C $\sqsubseteq$ \lbl{Sensitive})}
  \pline[2.4em]{$\lor$ ($\lnot$($\mathit{dest} \in$ BlockedDestinations)}
  \pline[3.0em]{$\land$ (Taint $\sqcup$ $\sigma$[$\mathit{src}$, content]).C $\sqsubseteq$ \lbl{Internal})}
  \pline[1.6em]{\kw{then} \kw{allow} \kw{else} \kw{deny} \pstr{"network upload exfiltration"}}%
}
\rblock{%
  \pline[0.8em]{\kw{rule} download\_integrity : \kw{on} downloadNet($\mathit{src}$, $\mathit{dest}$)}
  \pline[1.6em]{\kw{if} $\lnot$($\mathit{src} \in$ BlockedSources)}
  \pline[2.4em]{$\land$ ($\sigma$[$\mathit{src}$, content].I $\sqsubseteq$ $\sigma$[$\mathit{dest}$, content].I}
  \pline[3.4em]{$\lor$ $\sigma$[$\mathit{dest}$, content].C $\sqsubseteq$ \lbl{Internal})}
  \pline[1.6em]{\kw{then} \kw{allow} \kw{else} \kw{deny} \pstr{"untrusted or blocked download"}}%
}
\rblock{%
  \pline[0.8em]{\kw{rule} restrict\_blocked\_recv : \kw{on} recvNet($\mathit{src}$)}
  \pline[1.6em]{\kw{if} $\lnot$($\mathit{src} \in$ BlockedSources)}
  \pline[1.6em]{\kw{then} \kw{allow} \kw{else} \kw{deny} \pstr{"blocked network source"}}%
}
\rblock{%
  \pline[0.8em]{\kw{rule} restrict\_blocked\_concat : \kw{on} concatenate($\mathit{src}$, $\mathit{dest}$)}
  \pline[1.6em]{\kw{if} $\lnot$($\mathit{src} \in$ BlockedSources)}
  \pline[1.6em]{\kw{then} \kw{allow} \kw{else} \kw{deny} \pstr{"blocked concat source"}}%
}
\rblock{%
  \pline[0.8em]{\kw{rule} restrict\_undeclared\_send : \kw{on} sendNet($\mathit{dest}$)}
  \pline[1.6em]{\kw{if} $\mathit{dest} \in$ LocalMesh $\lor$ $\mathit{dest} \in$ AppServices $\lor$ $\mathit{dest} \in$ DeclaredHosts}
  \pline[1.6em]{\kw{then} \kw{allow} \kw{else} \kw{deny} \pstr{"undeclared send host"}}%
}
\rblock{%
  \pline[0.8em]{\kw{rule} restrict\_undeclared\_upload : \kw{on} uploadNet($\mathit{src}$, $\mathit{dest}$)}
  \pline[1.6em]{\kw{if} $\mathit{dest} \in$ LocalMesh $\lor$ $\mathit{dest} \in$ AppServices $\lor$ $\mathit{dest} \in$ DeclaredHosts}
  \pline[1.6em]{\kw{then} \kw{allow} \kw{else} \kw{deny} \pstr{"undeclared upload host"}}%
}
\rblock{%
  \pline[0.8em]{\kw{rule} restrict\_undeclared\_recv : \kw{on} recvNet($\mathit{src}$)}
  \pline[1.6em]{\kw{if} $\mathit{src} \in$ LocalMesh $\lor$ $\mathit{src} \in$ AppServices $\lor$ $\mathit{src} \in$ DeclaredHosts}
  \pline[1.6em]{\kw{then} \kw{allow} \kw{else} \kw{deny} \pstr{"undeclared receive host"}}%
}
\rblock{%
  \pline[0.8em]{\kw{rule} restrict\_undeclared\_download : \kw{on} downloadNet($\mathit{src}$, $\mathit{dest}$)}
  \pline[1.6em]{\kw{if} $\mathit{src} \in$ LocalMesh $\lor$ $\mathit{src} \in$ AppServices $\lor$ $\mathit{src} \in$ DeclaredHosts}
  \pline[1.6em]{\kw{then} \kw{allow} \kw{else} \kw{deny} \pstr{"undeclared download host"}}%
}
\rblock{%
  \pline[0.8em]{\kw{rule} restrict\_undeclared\_concat : \kw{on} concatenate($\mathit{src}$, $\mathit{dest}$)}
  \pline[1.6em]{\kw{if} $\mathit{src} \subseteq$ [LocalPaths, LocalMesh, AppServices, DeclaredHosts]}
  \pline[1.6em]{\kw{then} \kw{allow} \kw{else} \kw{deny} \pstr{"undeclared concat source"}}
  \vspace{2pt}
  \pline[0.8em]{\kw{default} \kw{allow}}%
}
\end{polbox}

\begin{polbox}[nobreak=true]{Privilege escalation policy}
\pline{\kw{security\_policy} privilegePolicy : Label \kw{where}}\nopagebreak
\rblock{%
  \pline[0.8em]{\kw{rule} block\_privilege\_escalation : \kw{on} escalatePrivilege($\mathit{target}$)}
  \pline[1.6em]{\kw{if} Taint $\sqsubseteq$ (\lbl{Internal}, \lbl{Trusted})}
  \pline[1.6em]{\kw{then} \kw{allow} \kw{else} \kw{deny} \pstr{"tainted privilege escalation"}}
  \vspace{2pt}
  \pline[0.8em]{\kw{default} \kw{allow}}%
}
\end{polbox}

\begin{polbox}[nobreak=true]{Combined benchmark policy}
\pline{\kw{security\_policy} benchmarkPolicy : Label \kw{where}}\nopagebreak
\pline[0.8em]{\kw{require} secretProtectionPolicy}
\pline[0.8em]{\kw{require} systemIntegrityPolicy}
\pline[0.8em]{\kw{require} antiExfiltrationPolicy}
\pline[0.8em]{\kw{require} privilegePolicy}
\pline[0.8em]{\kw{default} \kw{allow}}
\end{polbox}
\endgroup

\section{System-Call to Abstract-Action Mapping}
\label{App:syscall-mapping}

\tableref{event-to-action} summarizes how \isolate maps traced Linux system calls, network proxy records, and tracked process provenance ($\mathit{srcs}, \mathit{actor}, \mathit{origin}$) onto the abstract actions of~\sectref{dsl-transitions}.

\begin{table}[htb!]
  \centering
  \footnotesize
  \renewcommand{\arraystretch}{1.18}
  \caption{Mapping from traced system calls, proxy logs, and tracked provenance state ($\mathit{srcs}, \mathit{actor}, \mathit{origin}$) to the $11$ abstract actions.}
  \label{Ta:event-to-action}
  \begin{tabularx}{\textwidth}{@{}l >{\raggedright\arraybackslash}p{0.23\textwidth} >{\raggedright\arraybackslash}X@{}}
    \toprule
    \textbf{Abstract Action} & \textbf{Representative Syscalls / Sources} & \textbf{Emission Condition} \\
    \midrule
    $\act{read}(r, k)$ &
      \texttt{read}, \texttt{pread64}, \texttt{readv}, \texttt{mmap} (\texttt{PROT\_READ}), \texttt{getdents64}, \texttt{stat}/\texttt{fstatat}, \texttt{readlinkat}, \texttt{getxattr} &
      Reads file contents ($k = \pol{content}$) or metadata ($k \in \{\pol{location}, \pol{mode}, \pol{uid}, \pol{gid}, \pol{size}\}$) into the agent's context (either via \texttt{stdout}/\texttt{stderr} or via unconsumed process reads). \\[2pt]
    $\act{create}(\mathit{srcs}, r_{\text{dest}}, \mathit{actor}, \mathit{origin})$ &
      \texttt{openat} (\texttt{O\_CREAT}), \texttt{creat}, \texttt{mkdir}/\texttt{mkdirat}, \texttt{mknod} &
      Creates a path $r_{\text{dest}}$ that did not exist before the step; $\mathit{srcs}$ holds the files previously read by the process whose data may influence $r_{\text{dest}}$'s metadata. \\[2pt]
    $\act{write}(r_{\text{dest}}, k, \mathit{actor}, \mathit{origin})$ &
      \texttt{write}, \texttt{pwrite64}, \texttt{writev}, \texttt{truncate}, \texttt{mmap} (\texttt{MAP\_SHARED}), \texttt{chmod}, \texttt{chown}, \texttt{symlinkat} &
      Mutates $r_{\text{dest}}$ with no local source files ($\mathit{srcs} = \emptyset$): $k = \pol{content}$ for content writes/truncations, $k = \pol{mode}$ for \texttt{chmod}, $k \in \{\pol{uid}, \pol{gid}\}$ for \texttt{chown}, and $k = \pol{location}$ for \texttt{symlinkat}. \\[2pt]
    $\act{delete}(r_{\text{dest}}, \mathit{actor}, \mathit{origin})$ &
      \texttt{unlink}, \texttt{unlinkat}, \texttt{rmdir}, \texttt{rename}/\texttt{renameat2} &
      Removes $r_{\text{dest}}$, or removes the old source path after copying to the new path during a \texttt{rename}. \\[2pt]
    $\act{copy}(r_{\text{src}}, r_{\text{dest}}, \mathit{actor}, \mathit{origin})$ &
      \texttt{write}, \texttt{sendfile}, \texttt{splice}, \texttt{copy\_file\_range}, \texttt{mmap} (\texttt{MAP\_SHARED}), \texttt{rename}/\texttt{renameat2}, \texttt{linkat} &
      Overwrites or links $r_{\text{dest}}$ from a single source file $\mathit{srcs} = \{r_{\text{src}}\}$ (e.g., \texttt{cp}, \texttt{mv}, \texttt{ln}, or single-source shell redirection). \\[2pt]
    $\act{concatenate}(\mathit{srcs}, r_{\text{dest}}, \mathit{actor}, \mathit{origin})$ &
      \texttt{write}, \texttt{pwrite64}, \texttt{writev}, \texttt{copy\_file\_range}, \texttt{mmap} (\texttt{MAP\_SHARED}) &
      Overwrites $r_{\text{dest}}$ from multiple sources ($|\mathit{srcs}| \ge 2$), or appends/modifies $r_{\text{dest}}$ in place without truncation so $r_{\text{dest}}$'s prior content survives ($r_{\text{dest}} \in \mathit{srcs}$). \\[2pt]
    $\act{uploadNet}(r_{\text{src}}, r_{\text{net}})$ &
      \texttt{connect}, \texttt{sendto}, \texttt{sendmsg}, \texttt{write} + DNS / proxy log &
      Outbound network transmission to $r_{\text{net}}$ (emitted for both the resolved IP and each queried/claimed domain) when the sending process has read local files ($\mathit{srcs} \neq \emptyset$), once per $r_{\text{src}} \in \mathit{srcs}$. \\[2pt]
    $\act{sendNet}(r_{\text{net}})$ &
      \texttt{connect}, \texttt{sendto}, \texttt{sendmsg} + DNS / proxy log &
      Outbound network connection, DNS query, or request to $r_{\text{net}}$ when the sending process carries no local file sources ($\mathit{srcs} = \emptyset$). \\[2pt]
    $\act{downloadNet}(r_{\text{net}}, r_{\text{dest}})$ &
      \texttt{recvfrom}, \texttt{recvmsg}, \texttt{read} + file \texttt{write} &
      Network response from $r_{\text{net}}$ written to a local file $r_{\text{dest}}$ without entering \texttt{stdout}/\texttt{stderr}. \\[2pt]
    $\act{recvNet}(r_{\text{net}})$ &
      \texttt{recvfrom}, \texttt{recvmsg}, \texttt{read} + proxy log &
      Network response from $r_{\text{net}}$ written to \texttt{stdout}/\texttt{stderr} (or consumed into agent context without writing to a destination file). \\[2pt]
    $\act{escalatePrivilege}(p, b)$ &
      \texttt{execve} (\texttt{sudo}, \texttt{su}, \texttt{pkexec}, setuid), \texttt{setuid}, \texttt{setresuid} &
      Transitions process privilege to $p \in \{\lbl{User}, \lbl{Admin}\}$ via target binary or command $b$. \\
    \bottomrule
  \end{tabularx}
\end{table}

\end{document}